\documentclass[12pt, letterpaper, preprint]{aastex63}
\usepackage[T1]{fontenc}
\usepackage{fontawesome}
\usepackage{color}
\usepackage{amsmath}
\usepackage{natbib}
\usepackage{ctable}
\usepackage{bm}
\usepackage[normalem]{ulem} % Added by MS for \sout -> not required for final version
\usepackage{xspace}
\let\oldbibliography\thebibliography % killin' me.
\renewcommand{\thebibliography}[1]{%
  \oldbibliography{#1}%
  \setlength{\itemsep}{0pt}%
  \setlength{\parsep}{0pt}%
  \setlength{\parskip}{0pt}%
  \setlength{\bibsep}{0ex}
  \raggedright
}
\newcommand{\given}{\,|\,}

\newcommand{\bfi}[1]{\textbf{\textit{#1}}}

\newcommand{\lgal}{{\sc LGal}}
\newcommand{\fsps}{{\sc FSPS}}
\newcommand{\tlb}{t_{\rm lookback}}
\newcommand{\zmw}{Z_{\rm MW}}
\newcommand{\tage}{t_{\rm age, MW}}
\newcommand{\tauism}{\tau_{\rm ISM}}
\newcommand{\avgsfr}{\overline{\rm SFR}_{\rm 1Gyr}}
\newcommand{\avgssfr}{\overline{\rm SSFR}_{\rm 1Gyr}}
\let\oldAA\AA
\renewcommand{\AA}{\text{\normalfont\oldAA}}
\newcommand{\github}{\href{https://github.com/changhoonhahn/provabgs/}{\faGithub}}

\makeatletter \def\@captype{figure} \makeatother

\newcommand{\etal}{\emph{et~al.}}
\newcommand{\bitem}{\begin{itemize}}
\newcommand{\eitem}{\end{itemize}}
\definecolor{orange}{rgb}{1,0.5,0}

\shorttitle{PROVABGS Mock Challenge}
\shortauthors{Hahn et al.}

\begin{document} \sloppy\sloppypar\frenchspacing 

\title{The DESI PRObabilistic Value-Added Bright Galaxy Survey (PROVABGS) Mock Challenge} 
%\date{\texttt{DRAFT~---~\githash~---~\gitdate~---~NOT READY FOR DISTRIBUTION}}

\newcounter{affilcounter}
\author[0000-0003-1197-0902]{ChangHoon Hahn}
\correspondingauthor{ChangHoon Hahn}
\email{changhoon.hahn@princeton.edu}
\affiliation{Department of Astrophysical Sciences, Princeton University, Peyton Hall, Princeton NJ 08544, USA} 

\author[0000-0001-9802-362X]{K.J. Kwon}
\affiliation{Astronomy Department, University of California at Berkeley, Berkeley, CA 94720, USA}
\affiliation{Department of Physics, University of California, Santa Barbara, Santa Barbara, CA 93106-9530, USA}

\author{Rita Tojeiro}
\affiliation{School of Physics and Astronomy, University of St Andrews, North Haugh, St Andrews, KY16 9SS, UK}

\author{Malgorzata Siudek} 
\affiliation{Institut de F\'{\i}sica d'Altes Energies (IFAE), The Barcelona Institute of Science and Technology, 08193 Bellaterra, Barcelona, Spain}
\affiliation{National Centre for Nuclear Research, ul. Pasteura 7, 02-093, Warsaw, Poland}

\author{Rebecca E. A. Canning}
\affiliation{Institute of Cosmology \& Gravitation, University of Portsmouth, Dennis Sciama Building, Portsmouth, PO1 3FX, UK}

\author{Mar Mezcua}
\affiliation{Institute of Space Sciences (ICE, CSIC), Campus UAB, Carrer de Can Magrans, 08193, Barcelona, Spain}
\affiliation{Institut d’Estudis Espacials de Catalunya (IEEC), C/ Gran Capità, 08034 Barcelona, Spain}

\author{Jeremy L. Tinker}
\affiliation{Center for Cosmology and Particle Physics, Department of Physics, New York University, New York, USA, 10003}

%\suppressAffiliations
\author{David Brooks}
\affiliation{Department of Physics \& Astronomy, University College London, Gower Street, London WC1E 6BT, UK}

\author{Peter Doel}
\affiliation{Department of Physics \& Astronomy, University College London, Gower Street, London WC1E 6BT, UK}

\author{Kevin Fanning}
\affiliation{Department of Physics, The Ohio State University, 191 West Woodruff Avenue, Columbus, Ohio 43210, USA}

\author{Enrique Gazta\~{n}aga}
\affiliation{Institut de C\`{i}encies de l'Espai, IEEC-CSIC, Campus UAB, Carrer de Can Magrans s/n, 08913 Bellaterra, Barcelona, Spain}

\author{Robert Kehoe}
\affiliation{Department of Physics, Southern Methodist University, 3215 Daniel Avenue, Dallas, TX 75275, USA}

\author{Martin Landriau}
\affiliation{Lawrence Berkeley National Laboratory, 1 Cyclotron Road, Berkeley, CA 94720, USA }

\author{Aaron Meisner}
\affiliation{NSF's National Optical-Infrared Astronomy Research Laboratory, 950 N. Cherry Avenue, Tucson, AZ 85719, USA}

\author{John Moustakas}
\affiliation{Department of Physics and Astronomy, Siena College, 515 Loudon Road, Loudonville, NY 12211, USA}

\author{Claire Poppett}
\affiliation{Lawrence Berkeley National Laboratory, 1 Cyclotron Road, Berkeley, CA 94720, USA }

\author{Gregory Tarle}
\affiliation{Department of Physics, University of Michigan, Ann Arbor, MI 48109, USA}

\author{Benjamin Weiner} 
\affiliation{Department of Physics, University of Arizona, 1111 E. Fourth Street,
Tucson, AZ 85721}

\author{Hu Zou}
\affiliation{Key Laboratory of Optical Astronomy, National Astronomical
Observatories, Chinese Academy of Sciences, Beijing 100012, China}

\begin{abstract}
    The PRObabilistic Value-Added Bright Galaxy Survey (PROVABGS) catalog will
    provide measurements of galaxy properties, such as stellar mass ($M_*$),
    star formation rate (SFR), stellar metallicity ($Z_{\rm MW}$), and stellar
    age ($t_{\rm age, MW}$), for ${>}10$ million galaxies of the DESI Bright
    Galaxy Survey.
    Full posterior distributions of the galaxy properties will be inferred
    using state-of-the-art Bayesian spectral energy distribution (SED) modeling
    of DESI spectroscopy and Legacy Surveys photometry.
    In this work, we present the SED model, Bayesian inference framework, and
    methodology of PROVABGS. %\footnote{publicly available at }.
    Furthermore, we apply the PROVABGS SED modeling on realistic synthetic DESI
    spectra and photometry, constructed using the {\sc L-Galaxies}
    semi-analytic model.
    We compare the inferred galaxy properties to the true galaxy
    properties of the simulation using a hierarchical Bayesian framework
    to quantify accuracy and precision. 
    Overall, we accurately infer the true $M_*$, SFR, $Z_{\rm MW}$, and 
    $t_{\rm age, MW}$ of the simulated galaxies. 
    However, the priors on galaxy properties induced by the SED model have a
    significant impact on the posteriors.  %We characterize the priors and their impact in detail: 
    They impose a ${\rm SFR}{>}10^{-1} M_\odot/{\rm yr}$ lower bound on SFR, a
    ${\sim}0.3$ dex bias on $\log Z_{\rm MW}$ for galaxies with low spectral
    signal-to-noise, and $t_{\rm age, MW} < 8\,{\rm Gyr}$ upper bound on
    stellar age. 
    This work also demonstrates that a joint analysis of spectra and photometry
    significantly improves the constraints on galaxy properties over photometry
    alone and is necessary to mitigate the impact of the priors. 
    With the methodology presented and validated in this work, PROVABGS will
    maximize information extracted from DESI observations and provide a
    probabilistic value-added galaxy catalog that will extend current galaxy
    studies to new regimes and unlock cutting-edge probabilistic analyses.
    \github
\end{abstract}


% --- intro ---  
\section{Introduction} \label{sec:intro} 
% galaxy evolution with big data 
Large galaxy surveys have been transformational for our understanding of galaxy
evolution. 
With surveys such as the Sloan Digital Sky Survey~\citep[SDSS;][]{york2000},
Galaxy and Mass Assembly survey~\citep[GAMA;][]{driver2011}, and 
PRIsm MUlti-object Survey~\citep[PRIMUS;][]{coil2011}, 
we have now established the global trends of galaxies in the local universe. 
Population statistiscs, such as the stellar mass 
function~\citep{li2009, marchesini2009, moustakas2013} or quiescent
fraction~\citep{kauffmann2003a, blanton2003, baldry2006, taylor2009}, and their
evolution are now well understood. 
Many global scaling relations of galaxy propreties such as the mass-metallicity
relation~\citep{tremonti2004} or
the ``star formation sequence''~\citep{noeske2007, daddi2007,
salim2007} have
also been firmly established. 
Despite their importance in building our current understanding, however,
the empirical relations from existing observations are inadequate for
shedding further light on how galaxies form and evolve.

More precise and accurate measurements have the potential to reveal new trends
among galaxies undetected by previous observations.
So do new approaches that go beyond observed relations.
Empirical prescriptions for physical processes can be combined with $N$-body
simulations that capture hierarchical structure formation in empirical 
models~\citep[\emph{e.g.} {\sc UniverseMachine};][]{behroozi2019}. 
The predictions of these models can be compared to the observed distributions
of galaxy properties to derive insights into physical processes, such as the
timescale of star formation quenching~\citep{wetzel2013, hahn2017, tinker2017}. 
Predicted distributions of galaxy properties of large-scale cosmological 
hydrodyanmical simulations can also be compared to 
observations~\citep[\emph{e.g.}][]{genel2014, somerville2015a, dave2017a,
trayford2017, dickey2021, donnari2021}.
Though such comparisons are currently limited by the computation of costs of
simulations, advances in machine learning techniques for accelerating and
emulating simulations will enable such comparisons to explore a broad range of
galaxy formation models~\citep[\emph{e.g.}][]{villaescusa-navarro2021}.
Soon we will be able to compare detailed galaxy formation models directly
against observations and explore the parameter spaces and physical
prescriptions of the models. 
While many different approaches are available for expanding our understanding
of galaxies, they all require more statistically powerful galaxy samples with
well controlled systematics and well understood selection functions. 

Better observations, however, must be accompanied by better and more consistent
methodogy. 
The statistical power of large galaxy surveys are squandered when they are
analyzed inconsistently with a hodgepodge of methodologies since analyses
cannot take advantage of new techniques and approaches. 
In this regard, value-added catalogs (VACs) that provide consistently measured
galaxy properties for entire galaxy surveys are instrumental and have been
used by hundreds of galaxy studies~\citep[see][for a review]{blanton2009}.
For SDSS galaxies, the NYU-VAGC~\citep{blanton2005} provided photometric
properties (\emph{e.g.} absolute magnitudes) and the MPA-JHU
catalog~~\citep{brinchmann2004}\footnote{https://wwwmpa.mpa-garching.mpg.de/SDSS/DR7/}
provided spectral properties (\emph{e.g.} emission line luminosities).
Despite being released over a decade ago, these VACs are still widely used
today~\citep[\emph{e.g.}][]{alpaslan2021, odonnell2021, trevisan2021}. 

% motivate the probabilistic approach
Probabilistic VACs are the next advancement in VACs that will extract even more
the information from galaxy surveys. 
Unlike previous VACs that provide point estimates and rough estimates of
uncertainties, probabilistic catalogs provide full posterior distributions of
galaxy properties --- $p(\theta\given{\bfi X_i})$, the probability of galaxy
properties $\theta$ given observations, ${\bfi X_i}$, of galaxy $i$.
Posteriors offer more accurate measurements of galaxy properties because they
estimate the uncertainties and any degeneracies among them more accurately. 
They also open the doors for principled population inference.
Given observations of a set of galaxies, $\{{\bfi X_i}\}$, we can combine
individual posteriors of the galaxies to rigorously derive the distribution of
their physical properties: $p(\theta | \{ {\bfi X_i} \})$.
For example, from posteriors on stellar mass, $M_*$, and star formation
rate, SFR, we can infer $p(M_*, {\rm SFR} | \{ {\bfi X_i} \})$ by combining the
posteriors, or with the latest machine learning techniques~\citep{leja2021}.
This $M_*$-SFR distribution can then be used to measure the intrinsic width
star formation sequence with unprecedented accuracy and provide key insight
into star formation and stellar and AGN feedback in star-forming
galaxies~\citep[\emph{e.g.}][]{davies2021}.

With probabilistic catalogs, we can also include galaxies with less
tightly constrained properties in our analyses since posteriors accurately
quantify uncertainties. 
This means we can probe less explored, low signal-to-noise, regimes that may
shed new light on galaxy evolution, such as dwarf galaxies.
We can also more reliably quantify the fraction of extreme/outlier
galaxies, \emph{e.g.} quiescent fraction of field dwarf
galaxies~\citep{geha2012}.
Probabilistic catalogs also open the door for Bayesian Hierarchical approaches
and improve the statistical power of BGS through Bayesian shrinkage: the joint
posterior of the galaxy sample can be used as the prior to shrink the
uncertainties on the properties of individual galaxies. 
Overall, probabilistic VACs will enable a new level of statistical robustness
in galaxy studies and more fully extract the statistical power of galaxy surveys.

The PRObabilistic Value-Added Bright Galaxy Survey (PROVABGS) catalog will be a
probabilistic VAC constructed from the next pivotal large galaxy survey: the
Dark Energy Spectroscopic Instrument (DESI).
%In this work, we present the methodologies that will be used to construct the PRObabilistic Value-Added Bright Galaxy Survey (PROVABGS) catalog from the next pivotal large galaxy survey, the Dark Energy Spectroscopic Instrument (DESI) Bright Galaxy Survey (BGS).
%The Dark Energy Spectroscopic Instrument (DESI) marks the next stage in large galaxy surveys. 
Over the next five years, DESI will use its 5000 robotically-actuated fibers to
provide redshifts of ${\sim}30$ million galaxies over 
${\sim}14,000~{\rm deg}^2$, a third of the sky~\citep{desicollaboration2016,
desicollaboration2016a}.
The redshifts will be spectroscopically measured from optical spectra that
spans the wavelength range $3600 < \lambda < 9800\AA$ with spectral resolutions
$R = \lambda/\Delta \lambda = 2000 - 5000$.
In addition, DESI targets will also have photometry from the Legacy Imaging
Surveys Data Release 9~\citep[LS;][]{dey2019}, used for target selection. 
LS is a combination of three public projects (Dark Energy Camera Legacy Survey,
Beijing-Arizona Sky Survey, and Mayall $z$-band Legacy Survey) that jointly
imaged the DESI footprint in three optical bands ($g$, $r$, and $z$). 
It also includes photometry in the \emph{Wide-field Infrared Survey Explorer}
$W1$, $W2$, $W3$, and $W4$ infrared bands, derived from all imaging through
year 4 of NEOWISE-Reactivation force-photometered in the unWISE maps at the
locations of LS optical sources~\citep{meisner2017a, meisner2017}.

\begin{figure}
\begin{center}
\includegraphics[width=\textwidth]{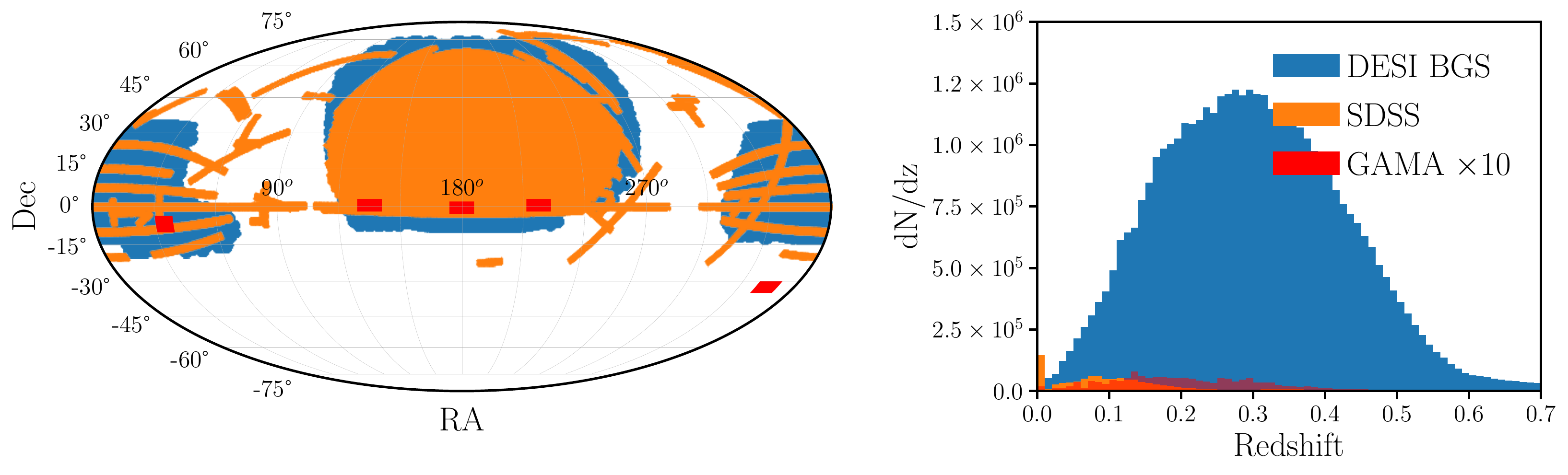} 
\caption{
    DESI will conduct the largest spectroscopic survey to date covering
    ${\sim}14,000~{\rm deg}^2$. 
    During dark time, DESI will measure ${>}20$ million spectra of luminous red
    galaxies, emission line galaxies, and quasars out to $z > 3$.
    During bright time, DESI will measure the spectra of ${\sim}10$ million
    galaxies out to $z{\sim}0.6$ with the Bright Galaxy Survey (BGS).
    {\em Left}: BGS (blue) will cover ${\sim}2\times$ the SDSS footprint
    (orange) and ${\sim}45\times$ the GAMA footprint (red).
    {\em Right}: We present the redshift distribution of BGS as predicted by
    the Millenium-XXL simulation~\citep[blue;][]{smith2017}. 
    We include the redshift distribution of SDSS and GAMA multiplied by
    $10\times$ for comparison. 
    BGS will be roughly two orders of magnitude deeper than the SDSS main
    galaxy sample and 0.375 mag deeper than GAMA.
    BGS will provide spectra for a magnitude limited sample of ${\sim}10$
    million galaxies down to $r < 19.5$ (BGS Bright) and a deeper sample of
    ${\sim 5}$ million galaxies as faint as $r < 20.175$ (BGS Faint).
}
\label{fig:bgs}
\end{center}
\end{figure}

During bright time, when the night sky is ${\sim}2.5\times$ brighter than
nominal dark conditions, DESI will conduct the Bright Galaxy Survey (BGS).
BGS will provide a $r < 19.5$ magnitude-limited sample of ${\sim}10$ million
galaxies out to redshift $z < 0.6$ --- the BGS Bright sample. 
It will also provide a surface brightness and color-selected sample of 
${\sim}5$ million faint galaxies with $19.5 < r < 20.175$ --- the BGS Faint
sample. 
The selection and completeness as well as the effect of systematics of the BGS
samples are characterized in detail in Hahn~\etal~(in prep.). 
Compared to the seminal SDSS main galaxy survey, BGS will provide optical
spectra two magnitudes deeper, over twice the sky, and double the median
redshift $z{\sim}0.2$ (Figure~\ref{fig:bgs}). 
It will observe a broader range of galaxies than previous surveys with
unprecendented statistical power. 
%BGS presents a unique opportunity to apply more sophisticated statistical analyses and new approaches to reveal the detailed connections among galaxy populations and  advance galaxy evolution studies. 

\begin{figure}
\begin{center}
    \includegraphics[width=0.5\textwidth]{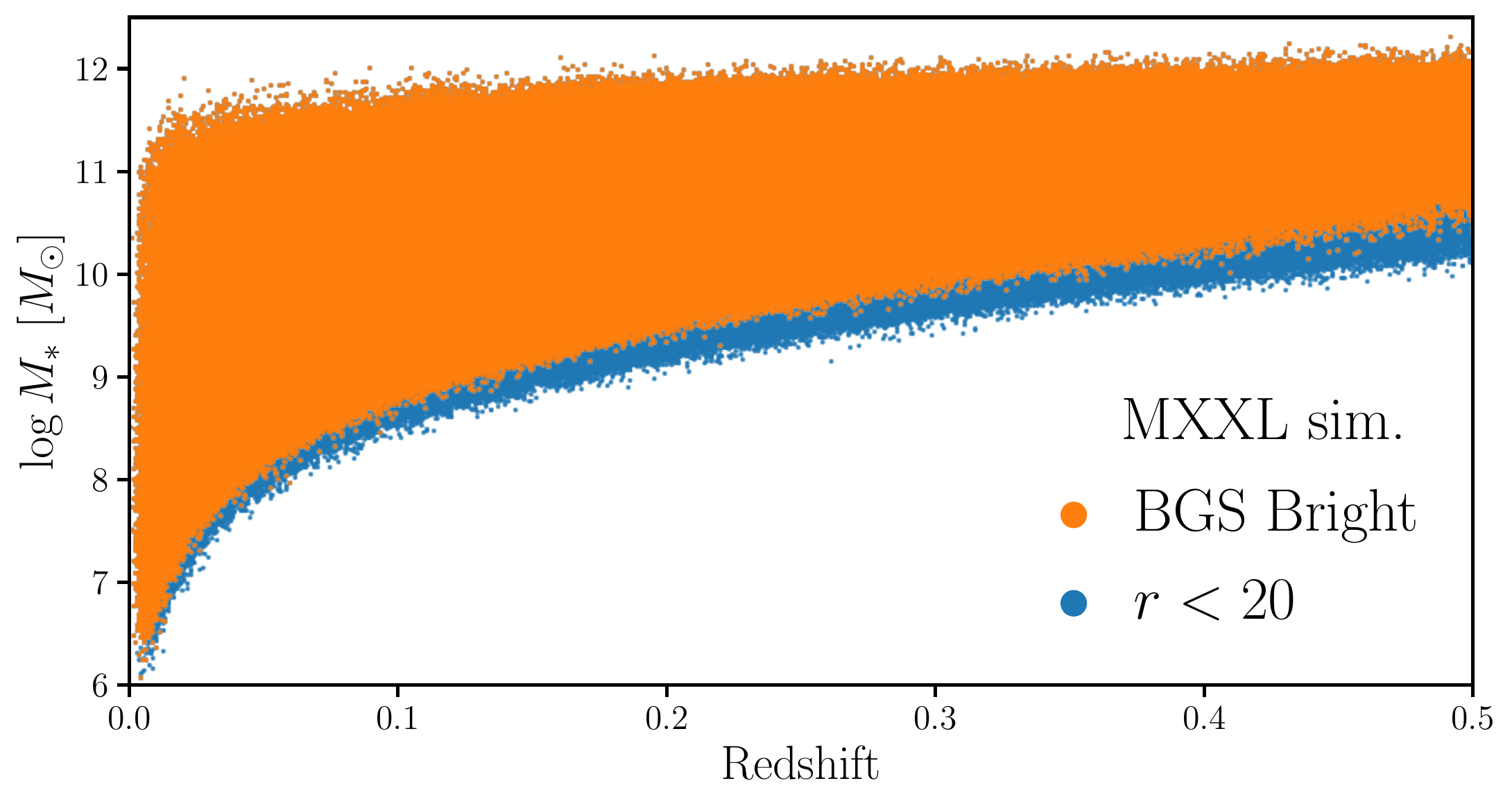}
    \caption{
        Stellar mass ($M_*$) distribution as a function of redshift of the 
        $r < 19.5$ magnitude-limited BGS Bright sample (orange) as predicted by
        the MXXL simulation. 
        We include the $M_*$ distribution of MXXL galaxies with $r < 20$ (blue)
        for reference.
        Many such fainter galaxies will be included in the BGS Faint sample,
        which will observe galaxies as faint as $r < 20.175$. 
        BGS will observe a broad range of galaxies with high completeness and
        provide galaxy samples with unprecedented statistical power.
        This includes a large sample of <$10^9M_\odot$ dwarf galaxies. %, which will produce tests of dark matter physics and galaxy physics.
        \emph{We will apply state-of-the-art Bayesian SED modeling to all BGS
        galaxies to construct the PRObabilistic Value-Added BGS (PROVABGS)
        catalog, which will unlock more sophisticated statistical approaches
        for galaxy evolution studies.}}\label{fig:bgs_mstar}
\end{center}
\end{figure}

%Better observations alone, however, are not sufficient.  
%The many advantages of BGS observations would be squandered if they were
%analyzed inconsistently with a hodgepodge of methodologies.  
%The advantages of new techniques and approaches cannot be realized without
%consistently measured galaxy properties. 
%Value-added catalogs (VACs) that provide consistently measured galaxy
%properties for entire galaxy surveys are instrumental in this regard and have
%been used by hundreds of galaxy studies~\citep[see][for a review]{blanton2009}.
%For SDSS galaxies, the NYU-VAGC~\citep{blanton2005} provided photometric
%properties (\emph{e.g.} absolute magnitudes) and the MPA-JHU
%catalog~~\citep{brinchmann2004}\footnote{https://wwwmpa.mpa-garching.mpg.de/SDSS/DR7/}
%provided spectral properties (\emph{e.g.} emission line luminosities).
%Despite being released over a decade ago, these VACs are still widely used
%today~\citep[\emph{e.g.}][]{alpaslan2021, odonnell2021, trevisan2021}. 
%Accompanying the BGS observations, the DESI Galaxy Quasar Physics (GQP) working
%group will produce the PRObabilistic Valued-Added Bright Galaxy Survey
%(PROVABGS) catalog. 

For all ${>}10$ million BGS galaxies, PROVABGS will provide full posterior
probability distributions of physical properties such as $M_*$, SFR,
metallicity ($Z$), and stellar age ($t_{\rm age}$). 
These properties will be inferred from both the LS photometry and DESI
spectroscopy using a state-of-the-art Bayesian modeling of the galaxy spectral
energy distribution (SED). 
PROVABGS will enable conventional analyses to be extended to a more
statistically powerful spectroscopic galaxy sample. 
Population statistics such as the stellar mass function or the star formation
sequence will be measured with higher precision than previously possible and
over a much wider range of galaxies (Figure~\ref{fig:bgs_mstar}). 
In particular, with the faint apparent magnitude limit of BGS ($r < 20.175$),
PROVABGS will include low mass (<$10^9M_\odot$) dwarf populations, which
provide important probes of the physics of dark matter and star formation
feedback.
The high completeness and simple selection function of the BGS Bright sample
will also facilitate comparisons to empirical models or galaxy formation
simulations with novel approaches. 

%Moreover, PROVABGS will provide more accurate measurements of galaxy properties
%from full posterior distributions, rather than point estimates. 
%Posteriors estimate the uncertainties on galaxy property measurements and any
%degeneracies among them more accurately. 
%This will enable a new level of statistical robustness in galaxy evolution
%studies.
%For instance, we can combine individual posteriors of galaxies to conduct
%principled population inference.
%Given observations of a set of galaxies, $\{X_i\}$, we can properly derive the
%distribution of their physical properties, $\theta$ --- $p(\theta | \{ X_i \})$
%--- and \emph{fully exploit the observations} 
%This also means that we can robustly include galaxies with less tightly
%constrained properties.  
%Little explored low signal-to-noise regimes that can be probed with this
%approach may shed new light on galaxy evolution.
%The PROVABGS posteriors will also open the door for Bayesian Hierarchical
%approaches and improve the statistical power of BGS through Bayesian shrinkage:
%the joint posterior of the galaxy sample can be used as the prior to shrink the
%uncertainties on the properties of individual galaxies. 
%With these advantages, PROVABGS will provide a VAC that fully exploits the
%DESI observations and maximizes the scientific impact of BGS. 

In this paper, we present the mock challenge for PROVABGS conducted by the DESI
Galaxy Quasar Physics working group. 
We present the state-of-the-art SED modeling that will be used to infer the
galaxy properties of BGS galaxies and construct the PROVABGS. 
We use an SED model with non-parametric prescriptions for galaxy star formation
and metallicity histories and accelerate the parameter inference using neural
emulators. 
Moreover, we validate our SED modeling on realistic mock BGS observations
constructed using the {\sc L-Galaxies} semi-analytic
model~\citep{henriques2015} and DESI survey simulations. 
By applying our SED model on mock observations, where we know the true galaxy
properties, we demonstrate that we can accurately infer galaxy properties for
PROVABGS and highlight the advantages of jointly analyzing photometry and
spectra. 
Furthermore, we characterize, in detail, the limits of our SED modeling so
that future studies using PROVABGS can use this work as a reference in
interpreting their results. 

In Section~\ref{sec:sims}, we describe the {\sc L-Galaxies} semi-analytic model
and how we use them to construct synthetic BGS observations. 
We then present the SED model, our Bayesian parameter inference framework with
neural emulators, and the mock challenge in Section~\ref{sec:methods}. 
We present the results of the mock challenge in Section~\ref{sec:results} and
discuss their implications in Section~\ref{sec:discuss}. 

% --- sims ---  
\section{Simulations}\label{sec:sims}
In this Section, we describe how we construct mock observations from simulated
galaxies of the {\sc L-Galaxies} semi-analytic galaxy formation model (SAM).
We use a forward model that includes realistic noise, instrumental effects, and
observational systematics to produce DESI-like photometry and spectra. 
Later, we apply Bayesian SED modeling to these mock observations and
demonstrate that we can accurately infer the true galaxy propertries.

\begin{figure}
\begin{center}
\includegraphics[width=\textwidth]{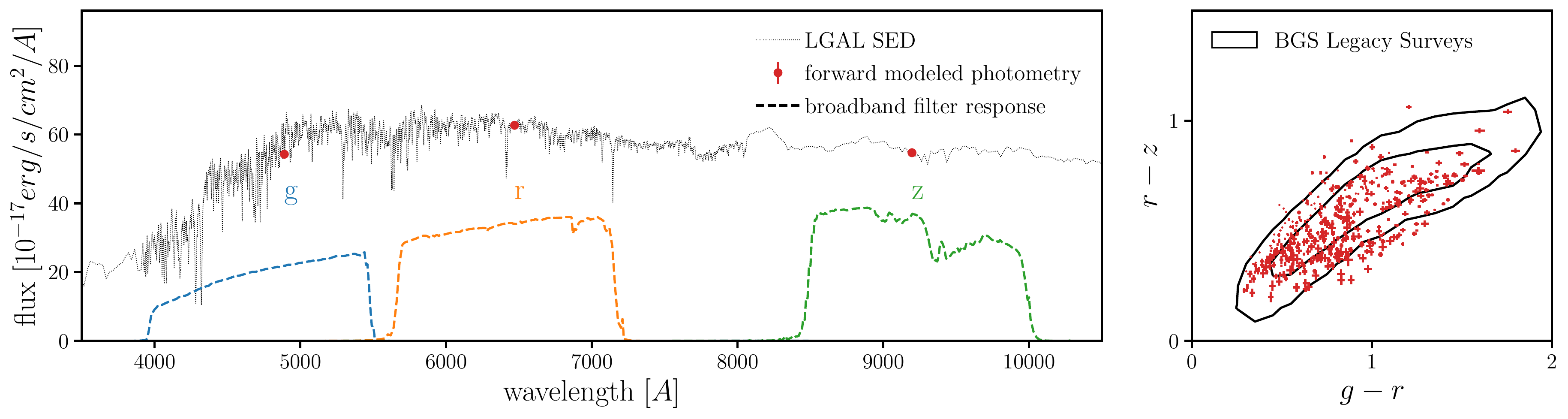}
\caption{{\em Left}: We forward model DESI optical $g$, $r$, and $z$ band
    photometry (red) for our simulated galaxies (Section~\ref{sec:lgal}) by
    convolving their SEDs (black dotted) with the broadband filters (dashed)
    and then applying an empirical noise model based on BGS objects in LS
    (Section~\ref{sec:photo}).
    {\em Right}: The $g-r$ and $r-z$ color distribution of the forward modeled
    \lgal~photometry is in good agreement with the color distribution of LS BGS
    objects (black contours). 
    We plot a subsample of the total 2,123 simulated galaxies from
    \lgal~that we use in this work.
    } \label{fig:photo}
\end{center}
\end{figure}

\subsection{L-Galaxies} \label{sec:lgal}
{\sc L-Galaxies}~\citep[hereafter \lgal;][]{henriques2015} is a state-of-the-at
semi-analytic galaxy formation model run on subhalo merger trees from the
Millennium~\citep{springel2005a} and Millenium-II~\citep{boylan-kolchin2009}
$N$-body simulations. 
Millenium-I and II provide a dynamic range of $10^{7.0} < M_* < 10^{12}
M_\odot$ and adopts a \cite{planckcollaboration2014a} $\Lambda$CDM cosmology.
\lgal~includes prescriptions for gas infall and cooling, star formation, disc
and bulge formation, stellar and black hole feedback, and the environmental
effects of tidal and ram-pressure stripping.
Feedback from active galactic nuclei (AGN), which prevents hot gas from
cooling, is the major mechanism for quenching star formation in massive
galaxies.
\lgal~model parameters are calibrated against the observed stellar mass
function and passive (quiescent) fraction at four different
redshifts from $z = 3$ to 0.
We refer readers to \cite{henriques2015} for further detail on \lgal.

% A recent comparison to the cosmological hydrodynamical simulation
% IllustrisTNG,  \lgal was recently compared to stellar mass functions and the stellar masses of individual galaxies agree
%  to better than  ∼0.2  dex with IllustrisTNG (Mohammadreza2021)

\subsection{Spectral Energy Distributions} \label{sec:sed}
For each simulated galaxy, \lgal~provides the star formation histories (SFHs)
and chemical enrichment histories (ZH) for its bulge and disk components,
separately, in approximately log-spaced lookback time bins.  
We treat each lookback time bin, $i$, as a single stellar population (SSP) of
age $t_i$.
We then derive the luminosities of the bulge and disk components by summing up
the luminosities of their SSPs:
\begin{equation}
    L^{\rm comp.}(\lambda) = \sum \limits_i \left({\rm SFH}^{\rm comp.}_i
    \Delta t_i\right)~L_{\rm SSP}(\lambda\,;\, t_i, Z^{\rm comp.}_i). 
\end{equation}
${\rm SFH}^{\rm comp.}_i$ and $Z^{\rm comp.}_i$ are the star formation rate and
metallicity of the bulge or disk component in lookback time bin $i$. 
$\Delta t_i$ is the width of the bin. 
$L_{\rm SSP}$ corresponds to the luminosity of the SSP, which we calculate
using the Flexible Stellar Population Synthesis~\citep[\fsps;][]{conroy2009,
conroy2010c} model.
% do we want to describe SPS models? 
For \fsps, we use the MIST isochrones~\citep{paxton2011, paxton2013,
paxton2015, choi2016, dotter2016} and the \cite{chabrier2003} initial mass
function (IMF). 
Also, we use the default spectral library in \fsps: the MILES spectral
library~\citep{sanchez-blazquez2006} over the wavelength range
$3800-7100\AA$ and the BaSeL
library~\citep{lejeune1997, lejeune1998, westera2002} 
outside of those limits.

% applying velocity dispersion to bulge and disk components separately 
Next, we apply velocity dispersions to $L^{\rm comp.}(\lambda)$.
For the disk, we apply a fixed $50\,\mathrm{km/s}$ velocity dispersion. 
For the bulge, we derive its velocity dispersion using the~\cite{zahid2016}
empirical relation that depends on the total bulge mass.
Afterwards, we apply dust attenuation to stellar emission in the disk component
($L^{\rm disk}$) based on the cold gas content and orientation of the disk. 
The attenuation curve is derived using a mixed-screen model with the
\cite{mathis1983} dust extinction curve. 
Stellar emission from stars younger than $30{\rm Myr}$ are further attenuated
with a uniform dust screen and a wavelength dependent optical depth.
No dust attenuation is applied to the bulge component.
We use the same dust attenuation that \cite{henriques2015} uses to construct
galaxy colors from \lgal~that match observations. 

Finally, we combine the attenuated disk component and the bulge component to
construct the total luminosity of the simulated galaxy and then convert this
rest-frame luminosity to observed-frame SED flux using its redshift, $z$.
\begin{equation}\label{eq:sed} 
    f_{\rm SED}(\lambda) = \frac{A(\lambda)L^{\rm disk}(\lambda) + L^{\rm bulge}(\lambda)}{4 \pi d_L(z)^2 (1+z)}.
\end{equation}
$A(\lambda)$ here is the dust attenuation for the disk component described
above and $d_L(z)$ is the luminosity distance.
In the left panel of Figure~\ref{fig:photo}, we present an example of the SED
flux constructed for an arbitrary \lgal~galaxy (black dotted).

\subsection{Forward Modeling DESI Photometry} \label{sec:photo} 
In this section, we describe how we construct realistic LS-like photometry
from the SEDs of simulated galaxies described in the last section.
First, we convolve the SEDs with the broadband filters of the LS to generate
broadband photometric fluxes: 
\begin{equation} \label{eq:photo}
    f_X = \int f_{\rm SED}(\lambda) R_X(\lambda) {\rm d}\lambda.
\end{equation}
$f_{\rm SED}$ is the galaxy SED (Eq.~\ref{eq:sed}) and $R_X$ is the
transmission curve for filter in the $X$ band. 
We generate photometry for the LS $g$, $r$, and $z$ optical bands.
Next, we apply realistic measurement uncertainties to the derived photometry by
sampling the noise distribution of BGS targets from LS DR9. 
We do this by matching each simulated galaxy to a BGS target with the nearest 
$r$-band magnitude and $g-r$ and $r-z$ colors.
The photometric uncertainties ($\sigma_X$) and $r$-band fiber flux ($f_r^{\rm
fiber}$) of the BGS object are then assigned to the simulated galaxy. 
We apply photometric noise by sampling a Gaussian distribution with standard
deviation $\sigma_X$: 
\begin{equation}
    \hat{f}_X = f_X + n_X  \quad {\rm where}~n_X \sim \mathcal{N}(0, \sigma_X).
\end{equation} 
Finally, we impose the target selection criteria of BGS~\citep[][Hahn~\etal~in
prep.]{ruiz-macias2021}.
In the left panel of Figure~\ref{fig:photo}, we overplot the forward
modeled photometry (red) on top of the SED flux (black) for an arbitrary
\lgal~galaxy. 
For reference, we also plot $R_X$ for the $g$, $r$, and $z$ bands of LS in
blue, orange, and green, respectively. 
On the right panel, we compare the $g - r$ versus $r - z$ color distribution
for the forward modeled \lgal~galaxies (red) to the color distribution of BGS
objects in LS (black contour). 
The errorbars represent the photometric uncertainties. 
The \lgal~galaxies have already been validated against observations, including
$UVJ$-band photometry~\cite{henriques2015}. 
However, we further confirm that the forward modeled photometry show good
agreement with LS BGS targets in optical color space.

\begin{figure}
\begin{center}
\includegraphics[width=0.8\textwidth]{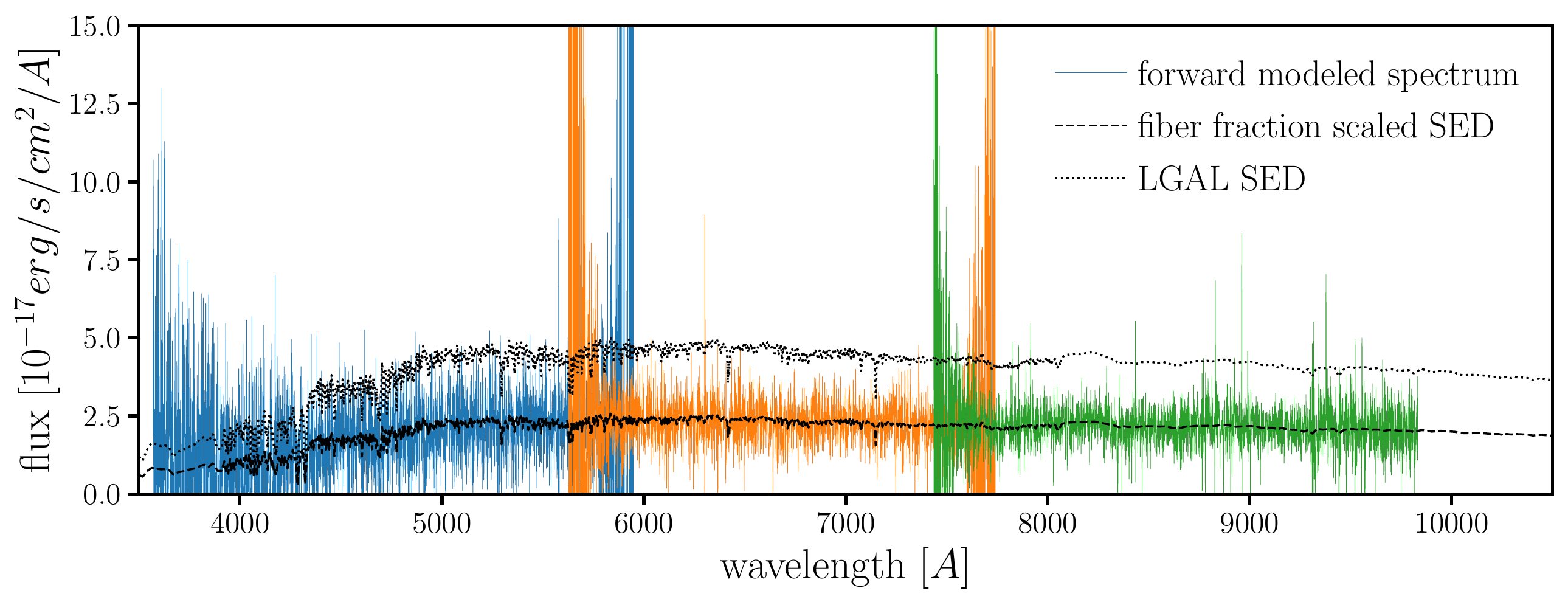}
\caption{
    We construct simulated DESI spectra (solid) for \lgal~simulated galaxies by
    applying a fiber aperture correction to the SED (dashed) and a realistic
    DESI noise model. 
    We apply a fiber aperture correction by scaling down the full SED (dotted)
    by the $r$-band fiber fraction derived from LS imaging. 
    The noise model accounts for the DESI spectrograph response and the bright
    time observing conditions of BGS (Hahn~\etal~in prep., Schlafly~\etal~in
    prep.).  
    We represent the spectra from the $b$, $r$, and $z$ arms of the DESI
    spectraphs in blue, orange, and green respectively. 
    Our forward model produces realistic DESI-like spectra that accurately
    reproduce the noise levels and characteristics of actual BGS spectra. 
    } \label{fig:spec}
\end{center}
\end{figure}

\subsection{Forward Modeling DESI Spectra} \label{sec:spec}
Next, we construct realistic DESI-like spectroscopy from the SEDs of simulated
galaxies. 
We begin by forward modeling the fiber aperture effect. % and apply a noise model that accurately reproduces the bright time observations of BGS. 
DESI uses fiber-fed spectrographs with fibers that have angular radii of 1''. 
Hence, only the light from a galaxy within this fiber aperture is collected by
the instrument.
Among BGS targets in LS, 40\% have $r_e$ < 1'' and 81\% have $r_e$ < 2'' so
the fiber aperture effect significantly impacts the majority of BGS
galaxies ($r_e$ is the half-light radius of the galaxy surface brightness
model fit by {\sc Tractor}\footnote{http://thetractor.org/doc/}).
To model this fiber apertuer effect, we use LS measurements of photometric
fiber flux within a 1'' radius aperture ($f_X^{\rm fiber}$), which estimates
the flux that passes through to the fibers.
When we assigned photometric uncertainties to our simulated galaxies based on
$r$, $g-r$, and $r-z$ in Section~\ref{sec:photo}, we also assigned $r$-band
fiber flux. 
We model the flux that passes through the fiber by scaling the SED flux by the
$r$ band fiber fraction, the ratio of $f_r^{\rm fiber}$ over the total $r$ band
flux: 
\begin{equation}
    f^{\rm spec}(\lambda) = \left(\frac{f_r^{\rm fiber}}{f_r}\right)f_{\rm SED}(\lambda).
\end{equation}
This fiber aperture correction assumes that there is no significant color
dependence. 
It also assume that there are no significant biases in the fiber flux
measurements in LS due to miscentering of objects. 
We discuss the implications of these assumptions later in
Section~\ref{sec:discuss} and will investigate them further in Ramos \etal~(in
prep.). 
In addition to the aperture correction, we also use $f_r^{\rm fiber}$ to derive
``measured'' $\hat{f}_r^{\rm fiber}$, since we do not know the true fiber
fraction in actual observations: 
\begin{equation}
    \hat{f}_r^{\rm fiber} = f_r^{\rm fiber} + n^{\rm fiber}_r \quad~{\rm
    where}~n^{\rm fiber}_r \sim \mathcal{N}\left(0, \frac{f_r^{\rm fiber}}{f_r}
    \sigma_r\right).
\end{equation}
We later use $\hat{f}_r^{\rm fiber}$ to set the prior on the nuisance parameter
of our SED modeling (Section~\ref{sec:methods}).

Next, we apply a noise model that simulates the DESI instrument response and
bright time observing conditions of BGS. 
We use the same noise model as the spectral 
simulations\footnote{\href{https://specsim.readthedocs.io/en/stable/guide.html}{https://specsim.readthedocs.io}} 
used for the BGS survey design and validation (Hahn~\etal~in prep.). 
We refer readers to Schlafly~\etal~(in prep.) for details about the survey
operations and simulations and Guy~\etal~(in prep.) for details on the DESI
spectroscopic data reduction pipeline.
Specifically, we use nominal dark time observing conditions with a $180s$
exposure time, which accurately reproduce the spectral noise and redshift
success rates of observed BGS spectra in early DESI observations.
In Figure~\ref{fig:spec}, we present the forward modeled BGS spectrum of an
arbitrary \lgal~galaxy (solid). 
We mark the spectrum from each arm of the three DESI spectrographs separately 
(blue, orange, green).
For reference, we include the full SED (dotted) and fiber fraction scaled SED
(dashed) of the galaxy.

%\todo{conclude by emphasizing the fact that our simulations cover the full expected
%observable space of DESI BGS and therefore if the pipeline works on our mocks,
%then it should work for every expected type of galaxies in the observations} 

% --- methods ---  
\newpage
\section{Joint SED modeling of Photometry and Spectra} \label{sec:methods}
\subsection{Stellar Population Synthesis Modeling} \label{sec:sps} 
PROVABGS will provide galaxy properties inferred from joint SED modeling of
DESI photometry and spectra. 
For the SED modeling, we use a state-of-the-art stellar population synthesis
(SPS) model that uses a non-parametric SFH with a starburst, a non-parametric
ZH that varies with time, and a flexible dust attenuation prescription. 

% describe SFH prescription
The form of the SFH is one of the most important factors in the accuracy of an
SPS model.
In general, the form of the SFH requires balancing between being flexible enough
to describe the wide range of SFHs in observations while not being too flexible
that it can describe any SFH at the expense of constraining power.  
If the model SFH is not flexible enough to describe actual SFHs of galaxies,
then unbiased galaxy properties cannot be inferred using the SPS model. 
For instance, most SPS models~(\emph{e.g.} CIGALE,~\citealt{serra2011,
boquien2019}; BAGPIPES,~\citealt{carnall2018}) use parametric SFH such as the
exponentially declining $\tau$-model.
Such functional forms, however, produce biased estimates of galaxy properties
(\emph{e.g.} $M_*$ and SFR) when used to fit mock observations of simulated 
galaxies~\citep{simha2014, pacifici2015, ciesla2017, carnall2018}.
On the other hand, many non-parametric forms of the SFH are overly flexible
and allow unphysical SFHs~\citep{leja2019}, which unncessarily increases 
parameter degeneracies and discards constraining power. 

\begin{figure}
\begin{center}
\includegraphics[width=0.85\textwidth]{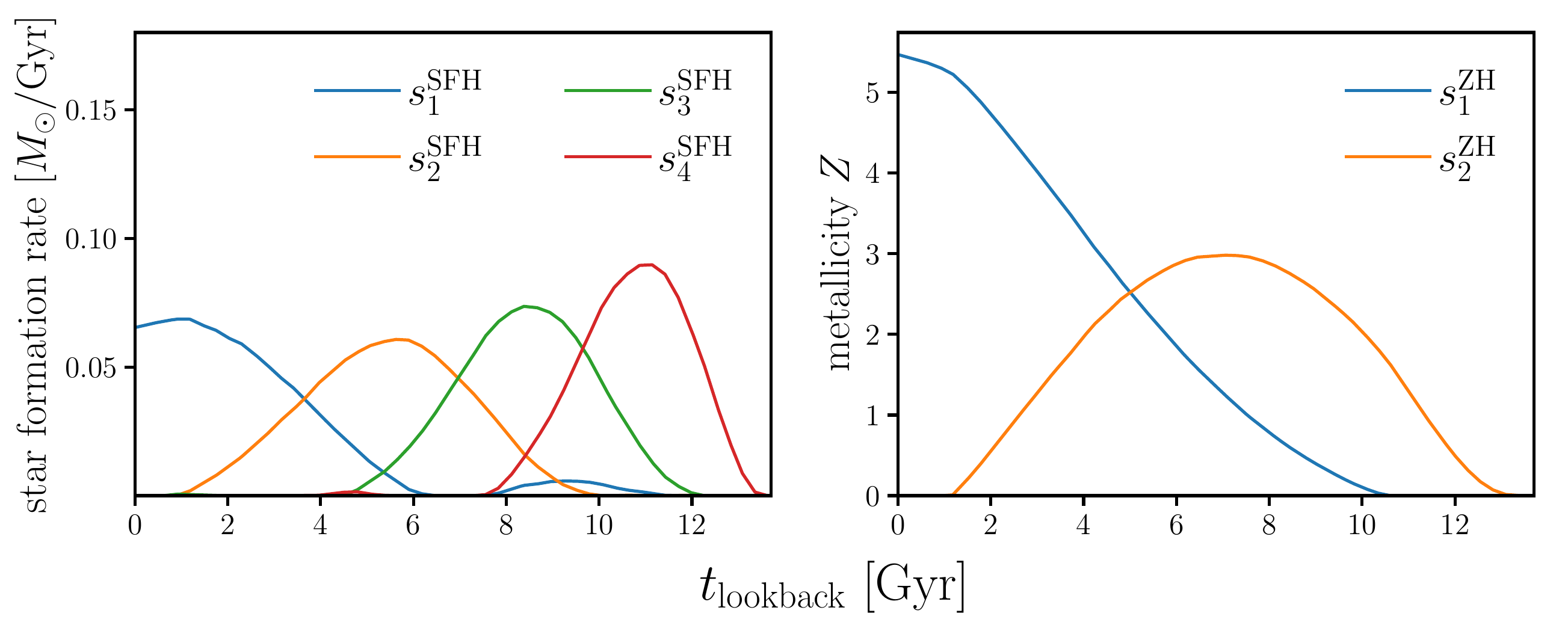} 
    \caption{
        Non-negative matrix factorization basis functions for the SFH (left)
        and ZH (right) used in the non-parametric SFH and ZH prescriptions of
        our SPS model. 
        These basis functions are derived from the SFHs and ZHs of simulated
        galaxies in the Illustris cosmological hydrodynamic simulations. 
        With the NMF basis functions, we can reproduce the wide range of SFHs
        and ZHs of Illustris galaxies (Appendix~\ref{sec:nmf}).  
    }
    \label{fig:nmf}
\end{center}
\end{figure}

In our SPS model, we use a non-parametric SFH with two components: one based on
non-negative matrix factorization~\citep[NMF;][]{lee1999,cichocki2009,
fevotte2011} basis functions and a starburst component.
For the first component, SFH is a linear combination of four NMF SFH bases:
\begin{equation} \label{eq:nmf} 
    {\rm SFH}^{\rm NMF} (t, t_{\rm age}) = \sum\limits_{i=1}^{4} \beta_i
    \frac{s_i^{\rm SFH}(t)}{\int\limits_0^{t_{\rm age}} s_i^{\rm SFH}(t) \,
    {\rm d}t}. 
\end{equation} 
$\{s^{\rm SFH}_i\}$ are the NMF basis functions and $\{\beta_i\}$ are the
coefficients. 
The integral in the denominator normalizes the NMF basis functions to unity. 
We constrain $\sum_i \beta_i = 1$, so the total SFH of the component over the
age of the galaxy ($t_{\rm age})$ is normalized to unity.
$\{s^{\rm SFH}_i\}$ are derived from the Illustris cosmological hydrodynamic
simulation~\citep{vogelsberger2014, genel2014, nelson2015}.
We compile, rebin, and smooth the SFHs of Illustris galaxies and then perform
NMF on them to derive $\{s^{\rm SFH}_i\}$. 
We find that 4 components is sufficient to accurately reconstruct the SFHs
from Illustris. 
We present the NMF SFH bases as a function of lookback time in
left panel of Figure~\ref{fig:nmf}.
By using NMF instead of \emph{e.g.} Principal Component Analysis (PCA), we
ensure that all of the SFH bases are non-negative and, thus, physically
meaningful. 
For further details on the derivation of the NMF bases, we refer readers to
Appendix~\ref{sec:nmf}. 
Assuming that the SFHs of Illustris galaxies resemble the SFHs of real
galaxies, our NMF form provides a compact and flexible representation of the
SFHs. 

The NMF basis functions are derived from smooth SFHs, which means that it does
not include any stochasticity. 
However, observations and high resolution zoom-in hydrodyanmical simulations
both find significant stochasticity in galaxy SFHs~\citep{sparre2017,
caplar2019, hahn2019b, iyer2020}. 
To include stochasticity in our SPS model, we include a starburst component
that consists of a SSP. 
Thus, for the total SFH, we use
\begin{equation} \label{eq:sfh}
    {\rm SFH} (t, t_{\rm age}) = (1 - f_{\rm burst})~{\rm SFH}^{\rm NMF} (t,
    t_{\rm age}) + f_{\rm burst}~\delta_{\rm D}(t - t_{\rm burst}).
\end{equation}
$f_{\rm burst}$ is the fraction of total stellar mass formed during the
starburst; $t_{\rm burst}$ is the time at which the starburst occurs; 
$\delta_{\rm D}$ is the Dirac delta function.
In total we use 6 free parameters in our SFH: 4 NMF basis coefficients 
($\beta_i$), $f_{\rm burst}$, and $t_{\rm burst}$. 

% describe ZH 
Another key part of an SPS model is the chemical enrichment history, or ZH. 
Current SPS models mostly assume a flat ZH, constant metallicity over
time~\citep{carnall2019a, leja2019}.
Since galaxies do not have constant metallicities throughout their history,
this assumption can significantly bias the inferred galaxy
properties~\citep{thorne2021}. 
Instead, we take a similar approach to the SFH and use NMF basis functions for
ZH:
\begin{equation}
    {\rm ZH}(t) = \sum\limits_{i=1}^2 \gamma_i s_i^{\rm ZH}(t).
\end{equation} 
$\{s_i^{\rm ZH}(t)\}$ are the ZH NMF basis functions and $\{\gamma_i\}$ are the
coefficients. 
$\{s_i^{\rm ZH}(t)\}$ are fit using the ZHs of simulated galaxies from
Illustris in the same fashion as the SFH. 
In the right panel of Figure~\ref{fig:nmf}, we present the ZH NMF bases as a
function of lookback time. 
We use two NMF components, so our ZH prescription has 2 free parameters. 

We use the SFH and ZH above to model the unattenuated rest-frame luminosity as
a linear combination of multiple SSPs, evaluated at logarithmically-spaced
lookback time bins.
We use a fixed log-binning with the bin egdes starting with $(0, 10^{6.05}{\rm
yr})$, $(10^{6.05}, 10^{6.15}{\rm yr})$, and continuing on with bins of width
0.1 dex.
The binning is truncated at the age of the model galaxy. 
For a $z=0$ galaxy, this binning produces 43 $\tlb$ bins.
We use log-spaced $\tlb$ bins because it better reproduces galaxy luminosities
evaluated with much higher resolution $\tlb$ binning than linearly-spacing, for
the same number of bins. 
At each of the 43 $\tlb$ bin $i$, we evaluate the luminosity of a SSP with
${\rm ZH}(t_i)$, where $t_i$ is the center of $\tlb$ bin, and total stellar
mass calculated by resampling the SFH in Eq.~\ref{eq:sfh}. 
We use \fsps~to evaluate the SSP luminosities and use the MIST isochrones, the
combination of MILES and BaSeL spectral libraries, and the \cite{chabrier2003}
IMF (same as in Section~\ref{sec:sed}).  
Since we use MIST isochrones, we impose a minimum and maximum limit to ${\rm
ZH}$ based on its coverage: $4.49\times10^{-5}$ and $4.49\times10^{-2}$,
respectively.
These metallicity values are in units of absolute metallicity and can be
converted to solar metallicity using $Z_\odot = 0.019$. 
We note that our stellar metallicity range is significantly broader than
previous studies for additional flexibility~\citep[\emph{e.g.}][]{leja2017,
carnall2019a, tacchella2021}. 
Since we model galaxies solely as a linear combination of SSPs, we do not
model nebular emission.  
We, therefore, exclude emission lines in our SED modeling by masking the
wavelength ranges of emission lines.

Before we combine the SSP luminosities, we apply dust attenuation.
We use a two component \cite{charlot2000} dust attenuation model with birth
cloud (BC) and diffuse-dust (ISM) components. 
The BC component represents the extra dust attenuation of young stars that are
embedded in modecular clouds and HII regions. 
For SSPs younger than $t_i < 100{\rm Myr}$, we apply the
following BC dust attenuation: 
\begin{equation}
    L_i(\lambda) = L_i^{\rm unatten.}(\lambda) \exp\left[-\tau_{\rm BC} \left(
    \frac{\lambda}{5500\AA} \right)^{-0.7} \right].
\end{equation}
$\tau_{\rm BC}$ is the BC optical depth that determines the strength of the BC
attenuation. 
Afterwards, {\em all} SSPs are attenuated by the diffuse dust using the
\cite{kriek2013} attenuation curve parameterization: 
\begin{equation}
    L_i(\lambda) = L_i^{\rm unatten.}(\lambda) \exp\left[-\tau_{\rm ISM} \left(
    \frac{\lambda}{5500\AA} \right)^{n_{\rm dust}} \left(k_{\rm Cal}(\lambda) +
    D(\lambda) \right) \right].
\end{equation}
$\tau_{\rm ISM}$ is the diffuse dust optical depth.
$n_{\rm dust}$ is the \cite{calzetti2001} dust index, which determines the
slope of the attenuation curve. 
$k_{\rm Cal}(\lambda)$ is the \cite{calzetti2001} attenuation curve and
$D(\lambda)$ is the UV dust bump, parameterized using a Lorentzian-like Drude 
profile:
\begin{equation}
    D(\lambda) = \frac{E_b(\lambda~\Delta \lambda)^2}{(\lambda^2 -
    \lambda_0^2)^2 + (\lambda~\Delta \lambda)^2}
\end{equation}
where $\lambda_0 = 2175 \AA$, $\Delta \lambda = 350\AA$, and 
$E_b=0.85 - 1.9\,n_{\rm dust}$ are the central wavelength, full width at half
maximum, and strength of the bump, respectively. 
Once dust attenuation is applied to the SSPs, we sum them up to get the
rest-frame luminosity of the galaxy. 
In total, our SPS model has 12 free parameters: $M_*$, 4 SFH basis
coefficients, $f_{\rm burst}$, $t_{\rm burst}$, 2 ZH basis coefficients,
$\tau_{\rm BC}$, $\tau_{\rm ISM}$, and $n_{\rm dust}$. 

%description of our speculator SED model \citep{alsing2019}, which is based on FSPS. We use Chabrier IMF \ch{do we need to justify htis?}. 

In practice, each model evaulation using \fsps~requires ${\sim}340$ ms. 
%evaluating each SSP using \fsps~requires \ch{X} seconds.  For each model evaluation, we evaluate $\sim 43$ SSPs in each of the log-spaced $\tlb$ bins. 
Though this is not a prohibitive computational cost on its own, sampling a
high dimensional parameter space for inference requires $>100,000$ evaluations
--- \emph{i.e.} $\gtrsim10$ CPU hours \emph{per galaxy}. 
For the >10 million BGS galaxies, this would require >100 million CPU hours. 
Instead, we use an emulator for the model luminosity, which uses a PCA neural
network (NN) following the approach of \cite{alsing2020}. 

%The NN provides a flexible and accurate mapping between the SPS model
%parameters and PCA coefficients --- \emph{i.e.} the NN predicts PCA
%coefficients for a given set of SPS parameters. 
%Then the linear combination of the predicted coefficients and PCA basis
%functions give us the emulated model luminosity. 
%The PCA basis functions and NN are trained using 1,000,000 SPS parameters and
%model luminosity pairs, $\{(\theta, L(\lambda;\theta))\}$. 

To construct our emulator, we first generate $N_{\rm model} = 1,000,000$
model luminosities, $L(\lambda;\theta)$, from unique SPS parameters,
$\theta$, sampled from the prior (Section~\ref{sec:infer},
Table~\ref{tab:params}).
We then split the model luminosities into four wavelength bins: 2000 - 3600,
3600 - 5500, 5500 - 7410, and 7410 - $60000\AA$ with $N_{\rm spec}$ = 127, 2109,
2113, and 549 resolution elements, respectively.
For each wavelength bin, a PCA is done in the $N_{\rm spec}$-dimensional
space to yield PCA basis functions, or eigenspectra. 
We represent the model luminosity using the first $N_{\rm basis}$ = 50, 50, 50, and 30
eigenspectra and their corresponding PCA coefficients. 
A NN is then trained on the set of $N_{\rm model}$ models to
derive a mapping from the 12 SPS parameters to the $N_{\rm basis}$ PCA
coefficients for each wavelength bin.

Once trained, our emulator works as follows.
For a given set of SPS parameters, the NN for each wavelength bin predicts
PCA coefficients. 
The coefficients are then linearly combined with the eigenspectra to
predict the model luminosity in the wavelength bin. 
The luminosity in all four wavelength bins are concatenated to produce the
full model luminosity. 
Throughout the wavelength range relevant for BGS, $3000 < \lambda < 9800\AA$,
we achieve $< 1\%$ accurate with the emulator. 
For details on the training, validation, and performance of our PCA NN
emulator, we refer readers to Kwon \etal~(in prep.). 
With the neural emulator, each model evaluation only requires 
${\sim}2.9$ ms --- 100$\times$ faster than with FSPS.

From the rest-frame luminosity, we obtain the observed-frame, redshifted, flux
in the same way as Eq.~\ref{eq:sed}.
In our case, redshift is not a free parameter since we will have high quality
spectroscopic redshifts for every DESI BGS galaxy.
BGS redshifts will have small redshift error, $\sigma_z < 0.0005 (1+z)$
(150 km/s), and <5\% catastrophic failures, $\Delta z/(1+z) < 0.003$ (<1000
km/s).
To model DESI photometry, we convolve the model flux with the LS broadband
filters as in Eq.~\ref{eq:photo}.
To model DESI spectra, we first apply Gaussian velocity dispersion. 
In this work, we keep velocity dispersion fixed at 0 km/s as a conservative
test for our SED modeling when we use an explicitly incorrect velocity
dispersion.
Later when we apply our SPS model to observations, the velocity dispersion will
be set to a more realistic value. 
It can also be set as a free parameter.
%In practice, however, the velocity dispersion can be set as a free parameter. 
After velocity dispersions, the broadened flux is resampled into the DESI
wavelength binning.  
Since DESI spectra do not necessarily include all the light of a galaxy, we
include a nuisance parameter $f_{\rm fiber}$, a normalization factor on the
spectra to account for fiber aperture effects. 
Next, the model photometry and spectrum can be directly compared to
observations.

\begin{table} 
\caption{Parameters of the PROVABGS SPS model and their priors used for joint
    SED modeling of DESI photometry and spectroscopy.} 
\begin{center}
    \begin{tabular}{ccc} \toprule
        name & description & prior \\[3pt]
        \hline 
        $\log M_*$                              & log galaxy stellar mass & uniform over [7, 12.5] \\
        $\beta_1, \beta_2, \beta_3, \beta_4$    & NMF basis coefficients for SFH & Dirichlet prior \\
        $f_{\rm burst}$ & fraction of total stellar mass formed in starburst event & uniform over [0, 1] \\
        $t_{\rm burst}$ & time of starburst event & uniform over [10Myr, 13.2Gyr] \\
        $\gamma_1, \gamma_2$ & NMF basis coefficients for ZH & log uniform over
        [$4.5\times10^{-5}, 1.5\times10^{-2}$] \\
        $\tau_{\rm BC}$ & Birth cloud optical depth & uniform over [$0, 3$] \\
        $\tau_{\rm ISM}$ & diffuse-dust optical depth & uniform over [$0, 3$] \\
        $n_{\rm dust}$ & \cite{calzetti2001} dust index & unifrom over[$-2, 1$]\\
        $f_{\rm fiber}$ & spectrum fiber-aperture effect normalization &
        Gaussian $\mathcal{N}(\hat{f}^{\rm fiber}_r, \frac{f^{\rm fiber}_r}{f_r} \sigma_r)$\\
        \hline            
\end{tabular} \label{tab:params}
\end{center}
\end{table}

\begin{figure}
\begin{center}
    \includegraphics[width=0.9\textwidth]{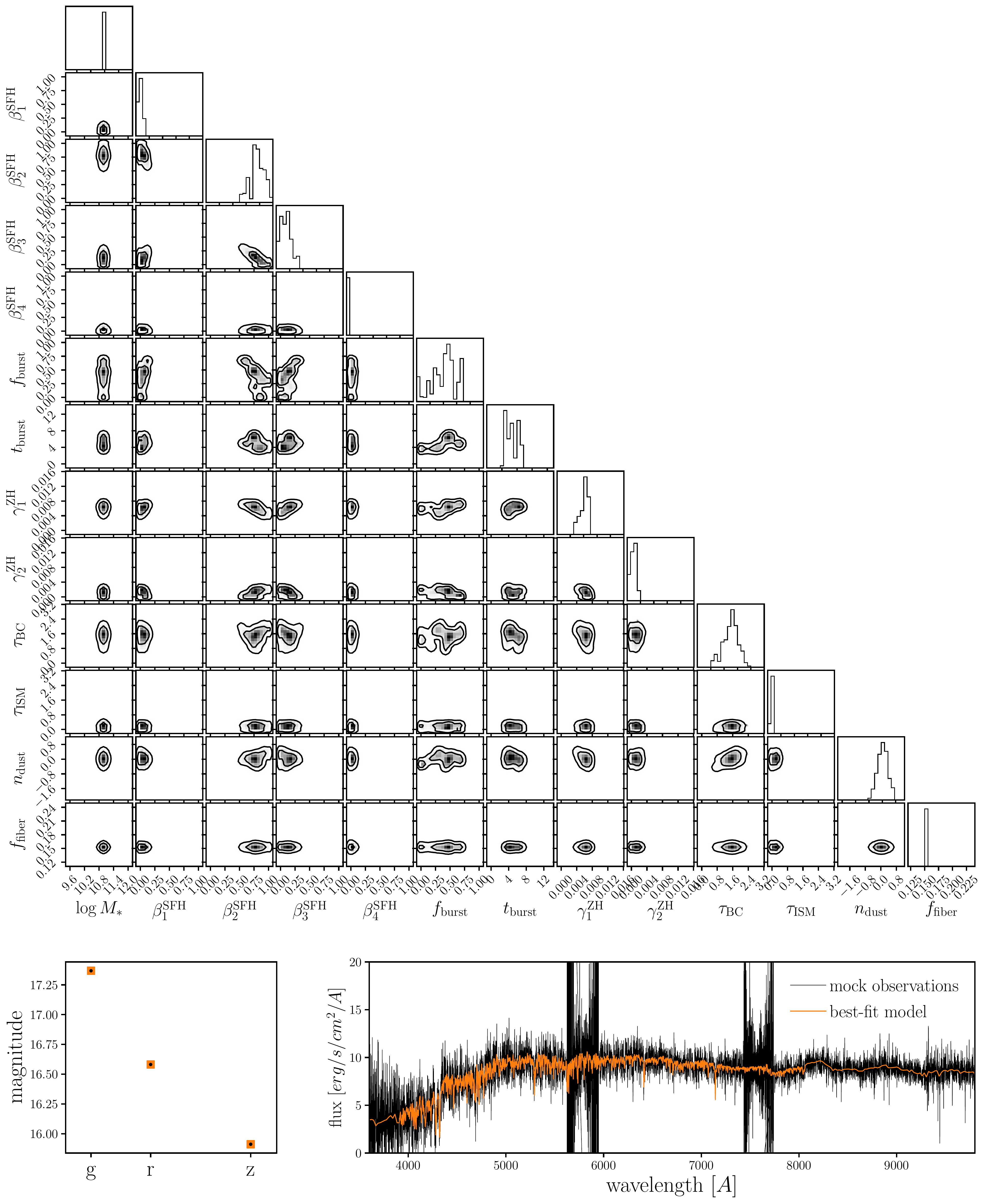}
    \caption{
        \emph{Top}: 
        Posterior probability distribution of our 12 SPS model parameters
        derived from joint SED modeling of the mock DESI photometry and
        spectrum.
        The contours mark the 68 and 95\% percentiles.
        We use a Gaussian likelihood and the prior specified in
        Table~\ref{tab:params} to evaluate the posterior and sample the
        distribution using ensemble slice MCMC. 
        \emph{With our Bayesian SED modeling approach, we accurately quantify
        uncertainties and capture complexities (\emph{e.g.}~parameter
        degeneracies and multimodality) in the posterior distribution.}\\
        \emph{Bottom}: 
        We compare the best-fit model observables (orange) to the mock
        observations (black).  
        We find excellent agreement for both the LS photometry (left) and the
        DESI spectrum (right). 
    } \label{fig:posterior}
\end{center}
\end{figure}

\subsection{Bayesian Parameter Inference} \label{sec:infer} 
Using the SPS model above, we perform Bayesian parameter inference to derive
posterior probability distributions of the SPS parameters from photometry and
spectroscopy. 
From Bayes rule, we write down the posterior as
\begin{equation} \label{eq:bayes}
    p(\theta\given {\bf X}) \propto p(\theta)~p({\bf X} \given \theta)
\end{equation}
where ${\bf X}$ is the photometry or spectrum and $\theta$ is the set of SPS
parameters. 
$p({\bf X} \given \theta)$ is the likelihood, which we calculate independently for
the photometry
\begin{equation}
    \mathcal{L}^{\rm photo} \propto \exp\left[-\frac{1}{2} \left(\frac{X^{\rm photo} -
    m^{\rm photo}(\theta)}{\sigma^{\rm photo}}\right)\right]
\end{equation}
and for the spectrum
\begin{equation}
    \mathcal{L}^{\rm spec} \propto \exp\left[-\frac{1}{2} \left(\frac{X^{\rm spec} -
    m^{\rm spec}(\theta)}{\sigma^{\rm spec}} \right)^2\right].
\end{equation}
$m^{\rm photo}$ and $m^{\rm spec}$ represent SPS model photometry and spectroscopy. 
$\sigma^{\rm photo}$ and $\sigma^{\rm spec}$ respresent the uncertainties on
the measured photometry and spectrum. 
In calculating $\mathcal{L}^{\rm spec}$, we exclude wavelength ranges of width
40\AA~surrounding the OII, H$\beta$, OIII, and H$\alpha$ emission lines since
our SED model does not model gas emissions.
We consider the photometry indepedent from the spectrum so we combine the
likelihoods when jointly modeling the spectrophotometry: 
\begin{equation}
    \log \mathcal{L} \approx \log \mathcal{L}^{\rm photo} + \log
    \mathcal{L}^{\rm spec}.
\end{equation}
$p(\theta)$ in Eq.~\ref{eq:bayes} is the prior on the SPS parameters. 
For most of our parameters, we use uninformative uniform priors with
conservatively chosen ranges that are listed in Table~\ref{tab:params}. 
However, for the priors of $\{\beta_1, \beta_2, \beta_3, \beta_4 \}$, the NMF coefficients
for the SFH, we use a Dirichlet distribution to maintain the normalization of
the SFH in Eq.~\ref{eq:nmf}. 
With Dirichlet priors, $\beta_i$ are within $0 < \beta_i < 1$ and
satisfy the constraint $\sum_i \beta_i = 1$. 

Now that we can evaluate the posterior at given $\theta$, we estimate the
posterior distributions using Markov Chain Monte Carlo (MCMC) sampling. 
We use the \cite{karamanis2020} ensemble slice sampling algorithm with the
{\sc zeus} Python
package\footnote{\href{https://zeus-mcmc.readthedocs.io/}{https://zeus-mcmc.readthedocs.io/}}. 
Ensemble slice sampling is an extension of standard slice sampling that does
not requires specifying the initial length scale or any further hand-tuning.
It generally converges faster than other MCMC algorithms (\emph{e.g.}
Metropolis) and generates chains with significantly lower autocorrelation.

When we sample the posterior, we do not directly sample our 12 dimensional
SPS parameter space because we use a Dirichlet prior on the SFH NMF
coefficients. 
Dirichlet distributions are difficult to directly sample so we instead use the
\cite{betancourt2012} sampling method, which transforms an $N$ dimensional
Dirichlet distribution into an easier to sample $N-1$ dimensional space.
Hence, we sample the posterior in the transformed 11 dimensional space. 
Given this dimensionality, we run our MCMC sampling with 30 walkers.
Overall, we find that the sampling converges after 2,500 iterations with a 500
iteration burn in. 
Deriving the posterior distribution from a joint SED modeling of photometry and
spectra, with the emulator, takes ${\sim}10$ CPU minutes per galaxy.
In principle, since our emulator uses a PCA NN, we can further expedite our
paremeter inference using more efficient sampling methods that exploit gradient
information, such as Hamiltonian Monte Carlo.  
We will explore further speed ups to our SED modeling in future works. 

In Figure~\ref{fig:posterior} we present the posterior distribution of our 12
SPS model parameters for an arbitrarily chosen \lgal~mock observation. 
We mark the 68 and 95 percentiles of the distribution with the contours. 
The posterior distribution reveal there are significant degeneracies between
SPS parameters: \emph{e.g.} $\beta_2^{\rm SFH}$ and $f_{\rm burst}$. 
Furthermore, the distribution is multimodal (see $f_{\rm burst}$ panels). 
With our Bayesian SED modeling, we are able to capture such complexities in the
posterior that would be lost with point estimates or maximum likelihood
approaches.
In the bottom panels, we compare our SPS model evaluated at the best-fit
parameters (orange) with the \lgal~mock observations (black). 
On the left, we compare the $g$, $r$, $z$ band magnitudes; on the right, we
compare spectra. 
We find excellent agreement between the best-fit SPS model and mock
observations.
The entire PROVABGS SED modeling pipeline, including the neural emulators and
parameter inference framework, is publicly available at
\href{https://github.com/changhoonhahn/provabgs/}{https://github.com/changhoonhahn/provabgs/}. 

% --- results ---  
\begin{figure}
\begin{center}
\includegraphics[width=0.95\textwidth]{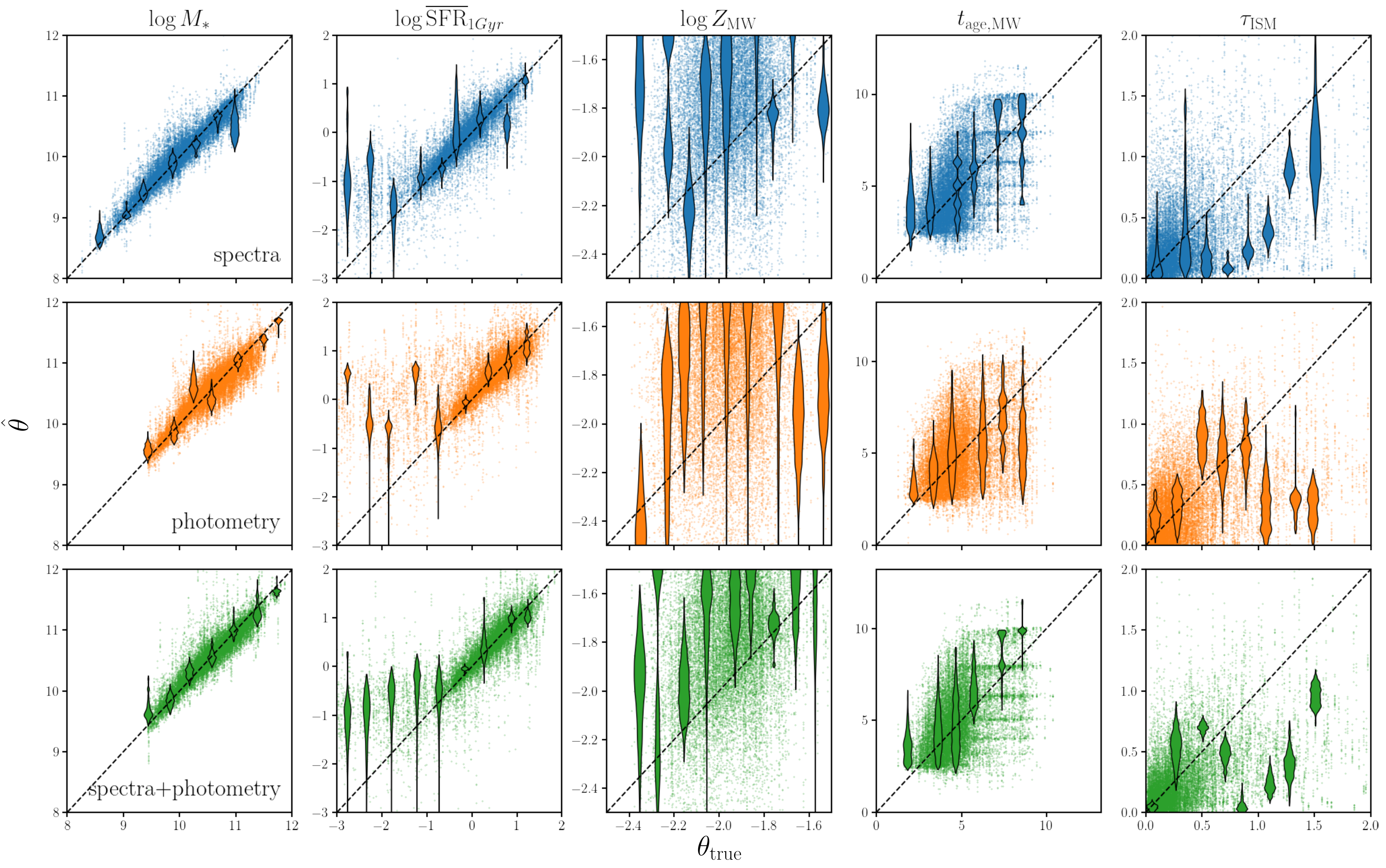}
\caption{
    Comparison between the true galaxy properties, $\theta_{\rm true}$, and
    those inferred from SED modeling of mock observations, $\hat{\theta}$. 
    From the left to right columns, we compare $\log M_*$, $\log \avgsfr$, 
    $\log Z_{\rm MW}$, $\tage$ and $\tauism$. 
    The inferred galaxy properties are derived from SED modeling of mock
    spectra (top), photometry (middle), and spectrophotometry (bottom). 
    For each simulated galaxy, we plot 10 samples drawn from the marginalized
    posterior of $\theta$.
    We also include violin plots, whose widths represent the marginalized
    posteriors, for a handful of randomly selected galaxies.
    \emph{The posteriors demonstrate that, overall, we can derive accurate and 
    precise constraints on certain galaxy properties from joint SED modeling of
    DESI photometry and spectra.}
    } \label{fig:prop_inf}
\end{center}
\end{figure}

\section{Results} \label{sec:results}
%\subsection{Inferred Galaxy Properties}
The goal of this work is to demonstrate the precision and accuracy of inferred
galaxy properties for PROVABGS. 
We apply our SED modeling to the mock observables of 2,123 \lgal~galaxies.
From the posterior distributions of the SPS parameters, we derive the following
physical galaxy properties: stellar mass ($M_*$), SFR averaged over 1 Gyr
($\avgsfr$), mass-weighted stellar metallicity ($Z_{\rm MW}$), mass-weighted
stellar age ($\tage$), and diffuse-dust optical depth ($\tau_{\rm ISM}$).
$M_*$ and $\tau_{\rm ISM}$ are SPS model parameters, while $\avgsfr$, 
$Z_{\rm MW}$, and $\tage$ are derived as 
\begin{equation} \label{eq:prop_eqs}
    \avgsfr = \frac{\int\limits_{t_{\rm age} - {\rm 1 Gyr}}^{t_{\rm age}}{\rm
    SFH}(t)\,{\rm d}t}{{\rm 1 Gyr}}, \quad
    Z_{\rm MW} = \frac{\int\limits_0^{t_{\rm age}}{\rm SFH}(t)\,{\rm
    ZH}(t)\,{\rm d}t}{M_*}, \quad{\rm and}\quad
    \tage = \frac{\int\limits_0^{t_{\rm age}}{\rm SFH}(t)\,t\,{\rm d}t}{M_*}.
\end{equation} 

In Figure~\ref{fig:prop_inf}, we compare the galaxy properties inferred from
SED modeling the mock observations, $\hat{\theta}$, to the true (input) galaxy
properties, $\theta_{\rm true}$, of the simulated galaxies.
From left to right, we compare $\log M_*$, $\log \avgsfr$, $\log Z_{\rm MW}$,
$\tage$, and $\tauism$ in each column.  
The inferred properties in the top, middle, and bottom rows are derived from
SED modeling of spectra, photometry, and spectrophotometry, respectively.
In each panel, we represent $\hat{\theta}$ by plotting 10 samples from the
marginalized posterior for each simulated galaxy. 
We also include violin plots of $\hat{\theta}$ for a handful of randomly
selected galaxies.
The width of the violin plot represents the marginalized posterior distribution
of $\theta$.
We note that in our SED modeling of spectra only, we do not include 
$f_{\rm fiber}$ so the true stellar mass in this case corresponds to 
$f_{\rm fiber} \times M_*$, which has a different range than for the 
photometry and spectrophotometry cases.
The comparison demonstrates that \emph{overall we robustly infer galaxy
properties using the {\sc PROVABGS} SED modeling}. 

In more detail, we find that we infer unbiased and precise constraints on
$M_*$ throughout the entire $M_*$ range. 
%For spectra+photometry SED modeling, the posteriors have $\sigma_{M_*}\sim0.06$ dex.
We also infer robust $\avgsfr$ above $\log \avgsfr > -1$ dex; below this limit,
however, the inferred $\avgsfr$ are significantly less precise and
overestimate the true $\avgsfr$. 
This bias at low $\avgsfr$ is caused by model priors, which we discuss in
further detail later in Section~\ref{sec:discuss} and
Appendix~\ref{sec:model_priors}. 
Both $Z_{\rm MW}$ and $\tage$ are not precisely constrained. 
The violin plots suggest that the inferred $Z_{\rm MW}$ overestimate the
true $Z_{\rm true}$.
For $\tage$, the posteriors are less precise for galaxies with older stellar
populations and they reveal the log-spaced $\tlb$ binning used in our
SPS model for $\tage > 6$ Gyr.
Lastly, $\tauism$ is overall accurately inferred for galaxies with low
$\tauism$ but appears to be underestimated for high $\tauism$.

The overall constraints on galaxy properties for the mock observations is
encouraging due to the significant differences in the forward model used to
generate the observations and the SPS model used in the SED modeling. 
First, the SFHs and ZHs in the mock observations are taken directly from
\lgal~simulation outputs while the SFH and ZH parameterization in the SPS model
is based on NMF bases fit to Illustris galaxies.
Second, in the forward model, we construct the SED of the bulge and disk
components of the simulated galaxies separately: the components have separate
SFHs and ZHs. 
The SPS model treats all galaxies as having one component. 
Third, we fix velocity disperions to 0 km/s in our SPS model. 
Lastly, we use different dust prescriptions: \cite{mathis1983} dust
attenuation curve in the forward model and the \cite{kriek2013} curve in the
SPS model. 
Despite these significant differences, our constraints on certain galaxy
properties are unbiased and precise. 

Figure~\ref{fig:prop_inf}, also highlights the advantages of jointly modeling
spectra and photometry. 
Comparing the constraints from spectrophotometry (bottom) versus photometry
alone (middle), we find that including spectra significantly tightens the
constraints for all properties. 
In addition, including spectra also appears to reduce biases of the
constraints. 
For instance, with only photometry, we derive significantly more biased
$\avgsfr$ constraints.
This is due to the limited constraining power of photometry, which allows the
posteriors to be dominated by model priors. 
Adding spectra, significantly increases the contribution of the likelihood and
ameliorates this effect. 
%Therefore, joint SED modeling of spectra and photometry 

Beyond qualitative comparisons of the posterior, we want to quantify the
precision and accuracy of the inferred galaxy properties. 
Let $\Delta_{\theta,i}$ be the discrepancy between the inferred and true
parameters for each galaxy: 
$\Delta_{\theta,i} = \hat{\theta}_i - \theta^{\rm true}_i$.
Then, if we assume that $\Delta_{\theta,i}$ are sampled from a Gaussian
distribution,
\begin{equation} \label{eq:eta_gauss}
    \Delta_{\theta,i} \sim \mathcal{N}(\mu_{\Delta_{\theta}}, \sigma_{\Delta_{\theta}}),
\end{equation}
the mean ($\mu_{\Delta_{\theta}}$) and standard deviation
($\sigma_{\Delta_{\theta}}$) of the distribution that represent the accuracy
and precision of the inferred posteriors for the galaxy population. 
We can infer the population hyperparameters, $\mu_{\Delta_{\theta}}$ and
$\sigma_{\Delta_{\theta}}$, using a hierarchical Bayesian
framework~\citep[\emph{e.g.}][]{hogg2010, foreman-mackey2014, baronchelli2020}.

Let $\{{\bfi X}_i\}$ represent the photometry or spectrum of a galaxy
population and $\eta_\Delta = \{\mu_{\Delta_{\theta}},
\sigma_{\Delta_{\theta}}\}$ represent the population hyperparameters.
Our goal is to constrain $\eta_\Delta$ from $\{{\bfi X}_i\}$ --- \emph{i.e.}
to infer $p(\eta_\Delta \given \{{\bfi X}_i\})$.
We expand 
\begin{align}\label{eq:popinf}
p(\eta_\Delta \given \{{\bfi X_i}\}) 
    =&~\frac{p(\eta_\Delta)~p( \{{\bfi X_i}\} \given \eta_\Delta)}{p(\{{\bfi X_i}\})}\\
    =&~\frac{p(\eta_\Delta)}{p(\{{\bfi X_i}\})}\int p(\{{\bfi X_i}\} \given \{\theta_i\})~p(\{\theta_i\} \given \eta_\Delta)~{\rm d}\{\theta_i\}.
\intertext{
    $\theta_i$ is the SPS parameters for galaxy $i$ and $p(\{{\bfi X_i}\}
    \given \{\theta_i\})$ is likelihood of the set of observations $\{{\bfi
    X_i}\}$ given the set of $\{\theta_i\}$. 
    Since the likelihoods for each of the $N$ galaxies, $p(\bfi X_i \given
    \theta_i)$, are not correlated, we can factorize and write the expression
    above as 
}
    =&~\frac{p(\eta_\Delta)}{p(\{{\bfi X_i}\})}\prod\limits_{i=1}^N\int p({\bfi X_i} \given \theta_i)~p(\theta_i \given \eta_\Delta)~{\rm d}\theta_i\\
    =&~\frac{p(\eta_\Delta)}{p(\{{\bfi X_i}\})}\prod\limits_{i=1}^N\int
    \frac{p(\theta_i \given {\bfi X_i})~p({\bfi X_i})}{p(\theta_i)}~p(\theta_i
    \given \eta_\Delta)~{\rm d}\theta_i\\
    =&~p(\eta_\Delta)\prod\limits_{i=1}^N\int \frac{p(\theta_i \given {\bfi
    X_i})~p(\theta_i \given \eta_\Delta)}{p(\theta_i)}~{\rm d}\theta_i. 
\intertext{
    $p(\theta_i \given {\bfi X_i})$ is the posterior for an individual galaxy,
    so the integral can be estimated using the Monte Carlo samples from the
    posterior: 
}
    \approx&~p(\eta_\Delta)\prod\limits_{i=1}^N\frac{1}{S_i}\sum\limits_{j=1}^{S_i}
    \frac{p(\theta_{i,j} \given \eta_\Delta)}{p(\theta_{i,j})}.
    \label{eq:popinf2}
\end{align} 
$S_i$ is the number of posterior samples and $\theta_{i,j}$ is the $j^{\rm th}$
sample of galaxy $i$.
$p(\theta_{i,j} \given \eta_\Delta) = p(\Delta_{\theta,i,j} \given
\eta_\Delta)$ is a Gaussian distribution and, hence, easy to evaluate. 
$p(\theta_{i,j}) = 1$ since we use uninformative and Dirichlet priors
(Table~\ref{tab:params}). 
Finally, we derive the maximum a posteriori (MAP) value of $\eta_\Delta$ by
maximizing the $p(\eta_\Delta \given \{{\bfi X_i}\})$ posterior distribution.
This type of population inference is a major advantage of inferring full
posteriors distributions of the galaxy properties.
We discuss the derivation and interpretation of the hyperparameters in more
detail in Appendix~\ref{sec:hyper}.

\begin{figure}
\begin{center}
    \includegraphics[width=0.9\textwidth]{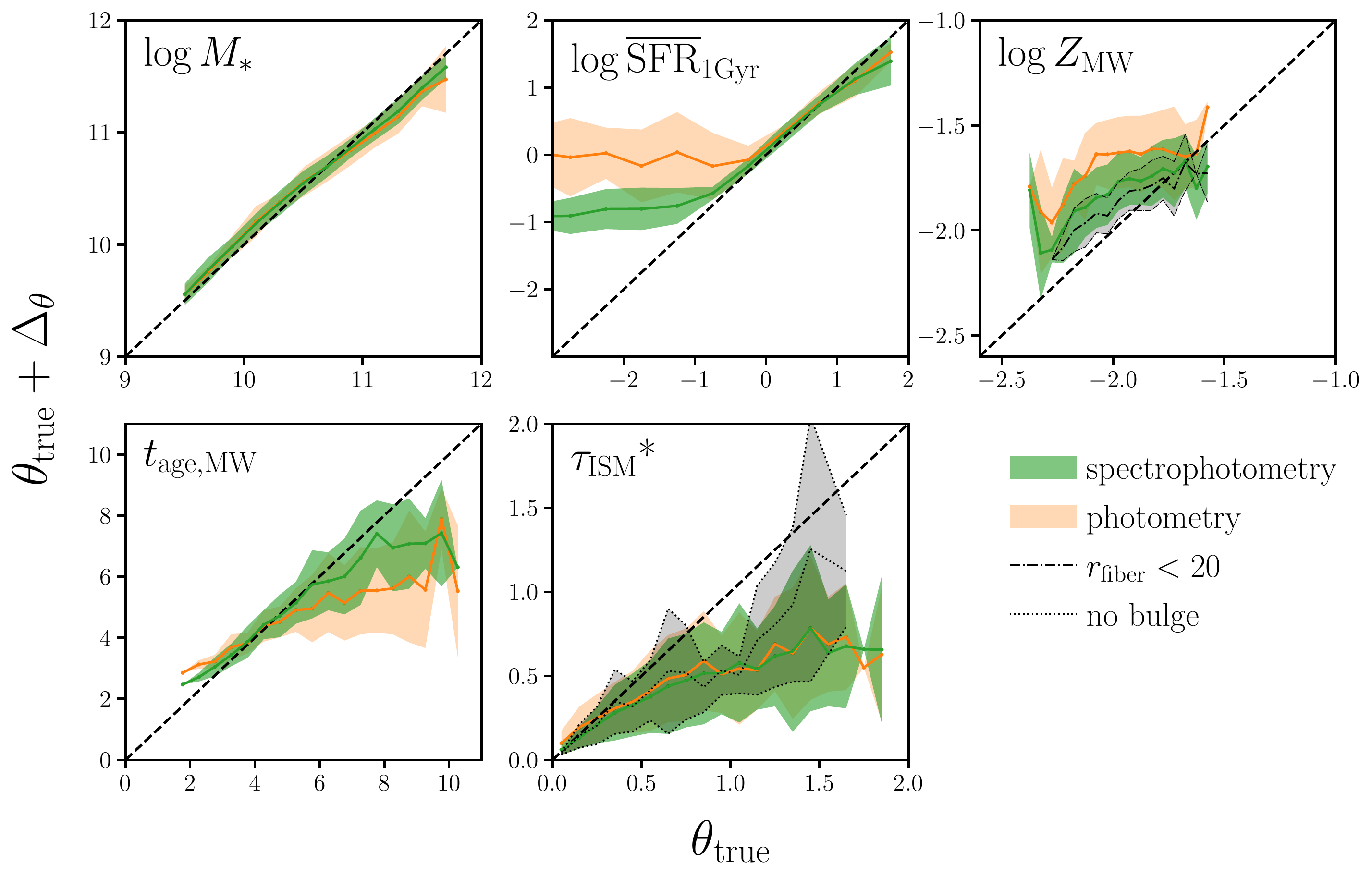}
    \caption{
        The accuracy and precision of galaxy property posteriors from our
        joint SED modeling of spectrophotometry, quantified using population
        hyperparameters $\eta_\Delta = \{\mu_{\Delta_{\theta}},
        \sigma_{\Delta_{\theta}}\}$, as a function of true galaxy property
        (green). 
        We derive $\eta_\Delta$ from the posteriors using a Hierarchical
        Bayesian approach. 
        We plot $\theta_{\rm true} + \mu_{\Delta_{\theta}}$ in solid line and
        represent $\sigma_{\Delta_{\theta}}$ with the shaded region.
        We include $\eta_\Delta$ for SED modeling of photometry alone (orange)
        for comparison. 
        Including DESI spectra significantly improves both the accuracy and
        precision of the inferred galaxy properties. 
        $\log\avgsfr$, $\log Z_{\rm MW}$, and $\tage$ constraints are
        significantly impacted by  priors imposed by the SPS model
        (Appendix~\ref{sec:model_priors}).
        Discrepancies in the dust prescriptions between our SPS model and the
        mock observations drive the bias in $\tauism$.
        Nevertheless, \emph{we accurately and precisely infer: $\log M_*$ for
        all $M_*$, $\log\avgsfr$ above $\log\avgsfr > -1\,{\rm dex}$, and 
        $\tage$ below $8\,{\rm Gyr}$.}
        } \label{fig:etas}
\end{center}
\end{figure}

In Figure~\ref{fig:etas}, we present the accuracy ($\mu_{\Delta_{\theta}}$) and
precision ($\sigma_{\Delta_{\theta}}$) of our joint SED modeling of spectra and
photometry (green) as a function of true galaxy property. 
$\mu_{\Delta_{\theta}}$ (solid) and $\sigma_{\Delta_{\theta}}$ (shaded region)
are the MAP values of $p(\eta_\Delta \given \{{\bfi X_i}\})$ posterior. 
In each panel, we derive $p(\eta_\Delta \given \{{\bfi X_i}\})$ for 
$\log M_*$, $\avgsfr$, $\log Z_{\rm MW}$, $\tage$, and $\tauism$ in bins of
widths 0.2 dex, 0.5 dex, 0.05 dex, 0.5 Gyr, and 0.1, respectively. 
We only include bins with more than ten galaxies. 
For comparison, we include $\eta_\Delta$ for SED modeling of photometry alone
(orange).
We also include $\eta_\Delta$ for $\log\zmw$ of galaxies with $r_{\rm fiber} >
20$ (black dot-dashed) and $\eta_\Delta$ for $\tauism$ of galaxies
without bulges (black dotted), which we discuss later. 

\begin{figure}
\begin{center}
    \includegraphics[width=\textwidth]{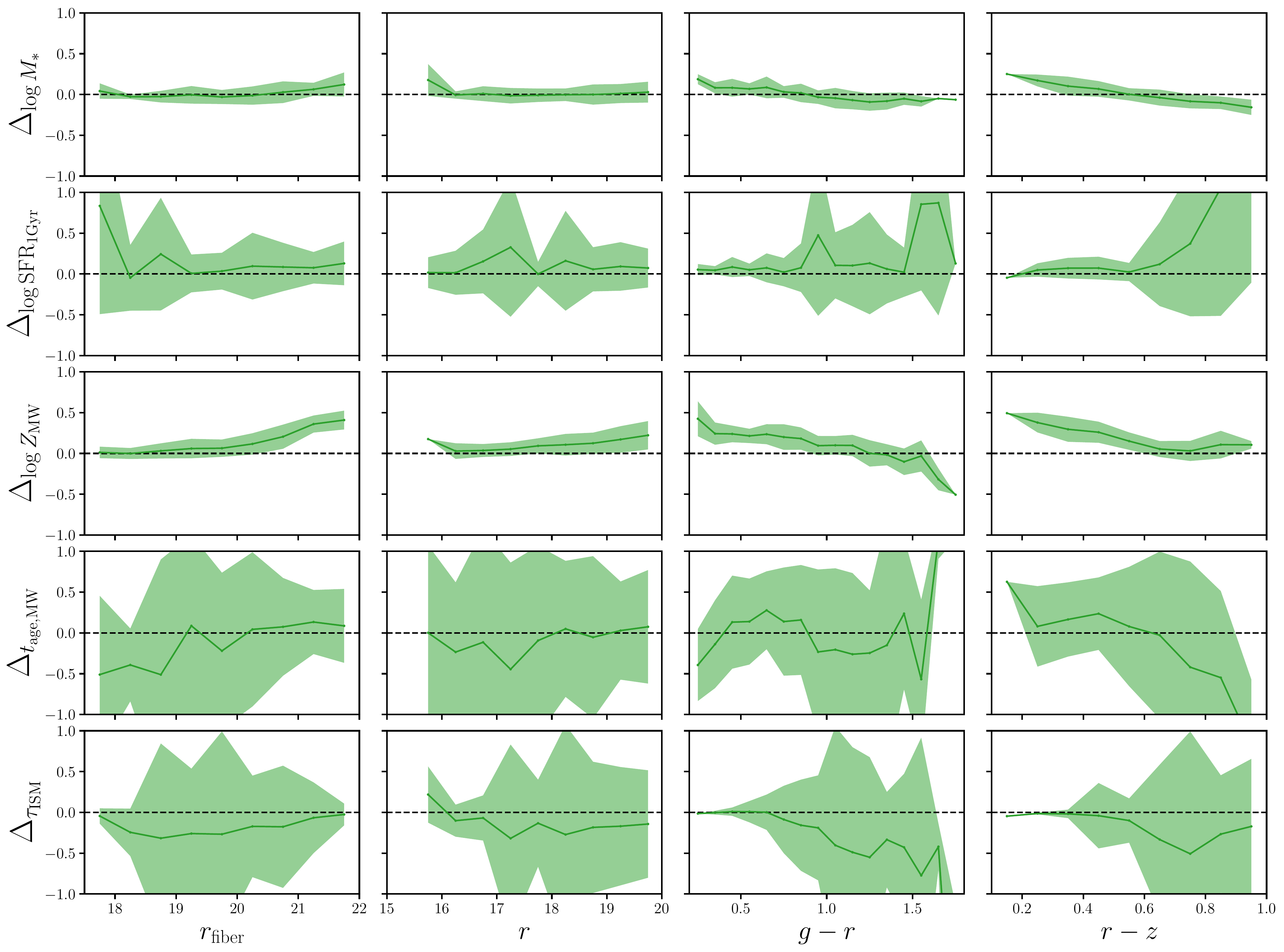}
    \caption{
        Accuracy and precision of the galaxy properties inferred from joint SED
        modeling of spectrophotometry as a function of $r_{\rm fiber}$, $r$,
        $g-r$, and $r-z$.
        $r_{\rm fiber}$ and $r$ magnitudes are proxies for spectral and
        photometric SNR. 
        From the top to bottom rows, we present $\eta_\Delta$ for $\log M_*$,
        $\log\avgsfr$, $\log Z_{\rm MW}$, $t_{\rm age, MW}$ and 
        $\tau_{\rm ISM}$.
        We find a significant dependence on spectral SNR in the inferred 
        $\log \zmw$. 
        When the spectral SNR is low ($r_{\rm fiber} > 20$), the prior on 
        $\log \zmw$ imposed by the SPS model dominate the posterior and
        cause $Z_{\rm MW}$ to be overestimated. 
        We find a significant color dependence on $\log\avgsfr$, $\log Z_{\rm
        MW}$, and $\tage$. 
        For $\log\zmw$ and $\tage$, the dependence is driven by underlying
        correlations with spectral SNR and true $\tage$. 
        Meanwhile, $\log\avgsfr$ is overestimated for the reddest galaxies with
        $r - z > 0.6$, which correspond to quiescent galaxies with $\log\avgsfr
        < -1$ dex. 
        Otherwise we find no significant dependence on SNR or optical color. 
    }    
    \label{fig:eta_photo}
\end{center}
\end{figure}

In Figure~\ref{fig:eta_photo}, we examine how the accuracy and precision of
our galaxy parameter constraints are impacted by signal-to-noise ratio (SNR) or
photometric color. 
We present $\eta_\Delta$ of our joint SED modeling of spectra and photometry as
a function of $r_{\rm fiber}$, $r$, $g-r$, and $r-z$. 
$r_{\rm fiber}$ and $r$ magnitudes serve as proxies of the SNR for the spectra
and photometry, respectively. 
In each row, we plot $\eta_\Delta$ for a different galaxy property: $\log M_*$,
$\avgsfr$, $\log Z_{\rm MW}$, $\tage$ and $\tauism$ (from top to bottom).

\begin{figure}
\begin{center}
    \includegraphics[width=\textwidth]{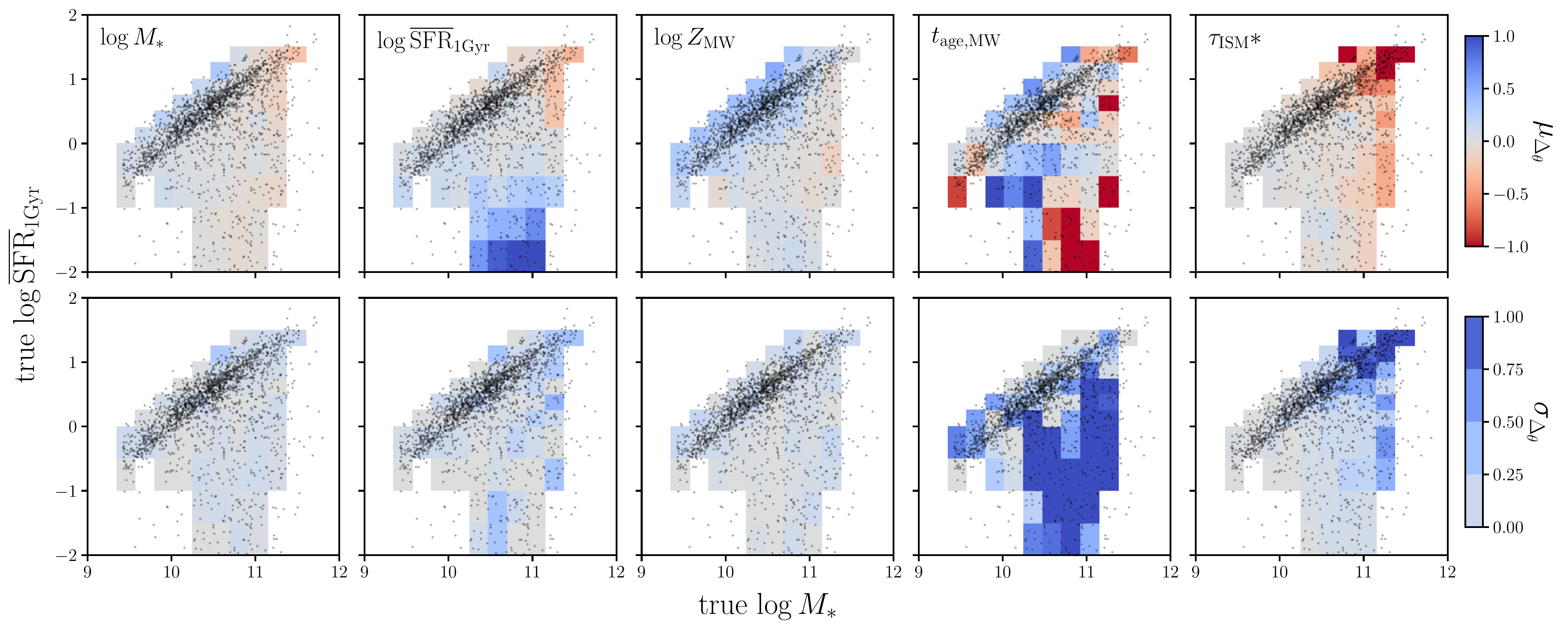} 
    \caption{
        Accuracy and precision of the galaxy properties inferred from joint SED
        modeling of spectrophotometry as a function of the galaxies' true $M_*$
        and $\avgsfr$. 
        We present $\mu_{\Delta_{\theta}}$ (top) and $\sigma_{\Delta_{\theta}}$
        (bottom) in ($M_*$, $\avgsfr$) bins for $\log M_*$, $\log\avgsfr$,
        $\log Z_{\rm MW}$, $t_{\rm age, MW}$ and $\tau_{\rm ISM}$ from left to
        right. 
        $\log M_*$ is accurately and precisely constrained for all types of galaxies. 
        $\log\avgsfr$ is accurately and precisely constrained for all galaxies
        except for quiescent galaxies with $\log\avgsfr < -1$ dex. 
        $\log\zmw$ is overestimated for star-forming galaxies, due to their
        overall lower spectral SNR. 
        $\tage$ is accurately and precisely constrained for star-forming
        galaxies that have overall younger stellar populations. 
        $\tauism$ is accurately and precisely constrained for all galaxies
        except massive star-forming galaxies, which have high true $\tauism$. 
    }\label{fig:etas_msfr}
\end{center}
\end{figure}

Lastly, in Figure~\ref{fig:etas_msfr}, we investigate whether there are any
underlying dependences in the inferred galaxy properties on the  
$M_*$-${\rm SFR}$ plane. 
In the top and bottom panels, we present $\mu_{\Delta_{\theta}}$ and 
$\sigma_{\Delta_{\theta}}$ in $(\log M_*, \log\avgsfr)$ bins for 
$\log M_*$, $\log\avgsfr$, $\log Z_{\rm MW}$, $\tage$ and $\tauism$ (left to
right).
We use $\log M_*$ bins of width 0.225 dex and $\log\avgsfr$ bins of width 
0.25 dex for $\log \avgsfr > 0$ dex and 0.5 dex for $\log \avgsfr < 0$ dex. 
We only show bins with more than 10 galaxies. 
On the $M_*-{\rm SFR}$ plane, we can examine whether the accuracy and precision
of the inferred properties have significant dependencies for galaxy type. 

Based on Figures~\ref{fig:etas}, \ref{fig:eta_photo}, and~\ref{fig:etas_msfr},
we draw the following conclusions on the accuracy and precision of the inferred
posteriors for each galaxy property:\\

% log M*
\noindent \underline{\emph{Inferred $\log M_*$}}: 
Overall, we infer accurate and precise $\log M_*$ from the {\sc PROVABGS} SED
modeling. 
There is no significant dependence in $\mu_{\Delta_{\theta}}$ and
$\sigma_{\Delta_{\theta}}$ with true $\log M_*$ throughout the $M_*$ range. 
We accurately infer the true $M_*$ throughout ${\sim}10^{9} - 10^{12} M_\odot$
with uniform precision of $\sigma_{\Delta_{\log M_*}}{\sim}0.1$ dex. 
We also find no significant dependence on SNR --- neither $r_{\rm
fiber}$ nor $r$ magnitudes significantly affect $\mu_{\Delta_{\log M_*}}$ and
$\sigma_{\Delta_{\log M_*}}$.
There is a noticeable correlation with $g-r$ and $r-z$ color, which also
appears in the $M_*-{\rm SFR}$ plane. 
However, this correlation is small compared to the precision of our inferred
posterior on $\log M_*$. 
When we compare the $\eta_\Delta$ from spectrophotometry to $\eta_\Delta$ from
photometry we find that including DESI spectra increases both the accuracy and
precision of the constraints, especially at high $M_* > 10^{11}M_\odot$. \\

% log SFR 
\noindent \underline{\emph{Inferred $\log\avgsfr$}}: 
We infer accurate $\log\avgsfr$ for galaxies with $\log\avgsfr > -1$ dex with
${\sim} 0.1$ dex precision. 
In fact, we find a $\log \avgsfr \sim -1$ dex lower bound for the inferred
$\log \avgsfr$.
Below this limit, we significantly overestimate $\log \avgsfr$, consistent with
the bias in Figure~\ref{fig:prop_inf}, and the constraints are also
significantly broader, $\sigma_{\Delta_{\log M_*}}{\sim}0.25 - 0.3$ dex.
Comparing $\mu_{\Delta_{\theta}}$ and $\sigma_{\Delta_{\theta}}$ from
spectrophotometry versus from only photometry, we confirm that including
spectra significantly improves the accuracy and tightens the $\log\avgsfr$
constraints.
For $\avgsfr$ below $\log\avgsfr < -1$ dex, including spectra reduces the bias
${\sim}1$ dex --- an order of magnitude. 

We find no significant correlation between the accuracy and precision of
$\avgsfr$ with spectral or photometric SNR.
However, there is a more significant color dependence where we overestimate
$\log\avgsfr$ by $\mu_{\Delta_{\log\avgsfr}}{>}0.5$ dex for the reddest
galaxies with $g-r > 1.5$ and $r-z> 0.6$.
The constraints for these galaxies are also significantly less precise:
$\sigma_{\Delta_{\log\avgsfr}} \sim 0.5$ dex. 
The bias is also apparent in Figure~\ref{fig:etas_msfr}, where we significantly
overestimate $\avgsfr$ for quiescent galaxies. 
$\avgsfr$ is also slightly underestimated for the most massive ($M_* >
10^{11}M_\odot$) star-forming galaxies. 
These biases are consequences of our SPS model priors.
$\avgsfr$ is a derived quantity; hence, the uninformative priors we impose on
SPS parameters induce non-uniform priors on them.
Our SPS model imposes a prior on $\log \overline{\rm SSFR}_{\rm 1 Gyr}$
that is skewed towards the peaks at $\sim$-10.4 dex
(Appendix~\ref{sec:model_priors}, Figure~\ref{fig:model_prior}). 
Consequently, the posterior overestimates $\avgsfr$ at low $\avgsfr$ (red,
quiescent galaxies) and underestimates $\avgsfr$ at the highest $\avgsfr$. \\

% log Z_MW  
\noindent \underline{\emph{Inferred $\log\zmw$}}:  
Unlike in Figure~\ref{fig:prop_inf}, $\eta_\Delta$ in Figure~\ref{fig:etas}
clearly reveals the accuracy and precision of the posteriors on $\log\zmw$. 
We find that $\mu_{\Delta_{\theta}}$ depends significantly on the true $\zmw$: 
inferred $\log\zmw$ is overestimated by ${\sim}0.2$ dex below $\log\zmw <
-2$ dex and slightly underestimated at the highest $\log\zmw > -1.6$ dex.
$\sigma_{\Delta_\theta} \sim 0.15$ dex is uniform throughout the $\zmw$ range.
Similar to $\avgsfr$, the bias in inferred $\zmw$ is a consequence of our SPS
model priors. 
The prior skews $\log\zmw$ constraints towards the peak of the prior at
$\log\zmw\sim-1.5$. 
Figure~\ref{fig:etas} also includes $\eta_\Delta$ for posteriors derived from
photometry alone (orange), which demonstrates that including DESI spectra
substantially improves the accuracy of the $\log\zmw$ constraints.
Including spectra reduces the overall bias on $\zmw$ by $\sim$0.3 dex. 
The improvement comes from the likelihood contribution from DESI spectra
reducing the relative contribution of the prior on the posterior. 

This is also why we find that the posteriors overestimate $\log\zmw$ at 
$r_{\rm fiber} > 20$ in Figure~\ref{fig:eta_photo}.
These correspond to mock observations with low spectral SNR where the
contribution of the likelihood from the spectra is lower and the prior on
$\log\zmw$ has a larger effect.
The color dependence of $\mu_{\Delta_\theta}$ for $\zmw$ in
Figure~\ref{fig:eta_photo} is also a consequence of this spectral SNR
dependence; so is the $M_*-{\rm SFR}$ dependence in Figure~\ref{fig:etas_msfr}.
If we exclude galaxies with low spectral SNR, both the color and $M_*-{\rm
SFR}$  dependences are substantially reduced: for $r_{\rm fiber} < 20$
galaxies, we infer $\log\zmw$ with $\mu_{\Delta_\theta}<0.15$ dex and
$\sigma_{\Delta_\theta}\sim0.1$ (Figure~\ref{fig:etas}; black dot-dashed). 
The $\zmw$ posteriors further underscore the constraining power of DESI
spectra. \\

% t_age, MW 
\noindent \underline{\emph{Inferred $\tage$}}:  
Figure~\ref{fig:etas} confirms that we derive unbiased and precise constraints
on $\tage$ out to $\tage < 8$ Gyr. 
Below this limit, we infer $\tage$ with $\sigma_{\Delta_\theta}{\sim}0.5$ Gyr.
For galaxies with older stellar populations above this limit, the log-spaced
$\tlb$ binning in our SPS model (Section~\ref{sec:sps}) expectedly
underestimates $\tage$ constraints and produces larger uncertainties 
($\sigma_{\Delta_{\tage}} \gtrsim 1$ Gyr). 
Meanwhile, we find no significant SNR or color dependence in
Figure~\ref{fig:eta_photo}. 
At $r - z > 0.6$, $\tage$ is underestimated, but this is driven by the
correlation between $r-z$ and true $\tage$: simulated galaxies with 
$r - z > 0.6$  have overall older stellar populations. 
In Figure~\ref{fig:etas_msfr}, we do not find a clear $M_*-{\rm SFR}$
dependence; however, $|\mu_{\Delta_{\tage}}|$ is larger and constraints are
significantly less precise for galaxies with older stellar populations below
the star-forming sequence.  \\

\noindent \underline{\emph{Inferred $\tauism$}}:  
Lastly, we find that both the accuracy and precision of our $\tauism$ depend
significantly on the true $\tauism$ value. 
The inferred constraints increasingly underestimate $\tauism$ with lower
precision for greater $\tauism$.
The bias is due to discrepancies between the dust prescriptions of SPS model
and the mock observations. 
First, we use a dust prescription with a different attenuation curve in the SPS
model than in the forward model. 
This places a strict limit on how accurately we can derive $\tauism$.
We intentially introduce this discrepancy since we do not know the ``true''
attenuation curve of observed galaxies in practice. 
Another reason for the biased $\tauism$ constraints is that we only attenuate
the stellar emission in the disk component of the simulated galaxies and not
the bulge component (Section~\ref{sec:sed}).
The true $\tauism$ is the optical depth for the disk component while our
$\tauism$ constraints correspond to the optical depth of dust attenuation
for the entire galaxies, a quantity that will be lower than the true $\tauism$
depending on how much the bulge contributes to the SED. 
Given these discrepancies, in this work we are primarily testing whether the
{\sc PROVABGS} SPS modeling can successfully marginalize over the effect of
dust and derive robust constraints on the other galaxy properties.

Nevertheless, we find no significant SNR or color dependence on the accuracy
and precision of $\tauism$ constraints (Figure~\ref{fig:eta_photo}). 
Furthermore, we find unbiased and precise $\tauism$ constraints for all galaxies
except star-forming galaxies above $M_* > 10^{11}M_\odot$ where we underestimate 
$\tauism$. 
Massive star-forming galaxies in this regime mainly have $\tauism > 1$.
In Figure~\ref{fig:etas}, we present a more apples-to-apples comparison of the
$\tauism$ constraints, where we present $\eta_\Delta$ for only galaxies without
bulge contributions (black dotted). 
For these galaxies, the bias in our $\tauism$ constraints is reduced and
$\mu_{\Delta_\theta}<0.5$ throughout the $\tauism$ range. 
Our constraints are still biased, however, due to the discrepant attenuation
curves. 
We emphasize that the primary goal of dust prescription in our SPS model is to
marginalize out the effect of dust. 
Based on the accuracy and precision of the constraints on other galaxy
properties, the {\sc PROVABGS} SPS model achieves this objective.  

% --- discussion ---  
\section{Discussion} \label{sec:discuss}
\subsection{Impact of Model Priors}
The most significant limitation of the {\sc PROVABGS} SED modeling in
inferring the true galaxy properties is the prior on galaxy properties imposed
by the model. 
The effect of such priors is a major limitation for any SED modeling
method~\citep[\emph{e.g.}][]{carnall2019a, leja2019} and is a consequence of
the fact that galaxy properties are \emph{not} parameters of the SPS model.
For instance, $\avgsfr$, $\zmw$, and $\tage$ are derived by integrating the SFH
and ZH (Eq.~\ref{eq:prop_eqs}), which are parameterized by $\beta_1, \beta_2,
\beta_3, \beta_4$, $f_{\rm burst}$, $t_{\rm burst}$, and $\gamma_1, \gamma_2$. 
The uniform and Dirichlet priors on these parameters (Section~\ref{sec:infer}
and Table~\ref{tab:params}) do not translate into uniform priors on $\avgsfr$,
$\zmw$ and $\tage$.
Other galaxy properties (\emph{e.g.}~SFH, and ZH) likewise have
non-uniform, and undesireable, priors. 

One way to address this issue is to choose an SED model parameterization that
does not impose extreme priors on galaxy properties and to characterize the
priors in detail so that final posteriors can be appropriately interpreted. 
For the {\sc PROVABGS} model, we explicitly chose our SFH prescription so that
the prior on $\log\avgssfr$ spans the range $-12$ to $-9$ dex.
Furthermore, we fully characterize the prior on $\avgssfr$, $\zmw$, $\tage$,
SFH, and ZH in Appendix~\ref{sec:model_priors} (Figures~\ref{fig:model_prior}
and~\ref{fig:sfh_prior}). 
This way, we understand exactly how the model prior impacts the derived
posteriors as we discuss in Section~\ref{sec:results}. 
Beyond mitigating the effect of the priors, we can alternatively impose uniform
prior (or any other desired prior distribution) on the derived galaxy
properties by adjusting the priors on the SED model parameters. 
\cite{handley2019} recently demonstrated that maximum-entropy priors can be
used for this purpose to impose uniform priors on the inferred sum of neutrino
masses in cosmological analyses. 
In an upcoming paper, Hahn (in prep.), I will demonstrate that maximum-entropy
priors can also be used in Bayesian SED modeling to correct for the impact of
priors on infer posteriors on derived galaxy properties. 

%With this prior correction, we will be able to infer even more accurate posteriors on the physical properties of galaxies with our {\sc PROVABGS} SED modeling.

%However, we can go beyond minimizing the impact of the priors and use maximum-entropy.  With an estimate of the prior distribution, we can impose maximum-entropy priors in a specified distribution~\citep{handley2019}.  From an estimate of the prior distribution on the galaxy properties, we can derive a new prior on the SPS model parameters that would impose uniform priors on the galaxy properties. 

\subsection{Aperture Effects}
In this work, we use forward modeled mock observations to demonstrate that we
can infer accurate and precise posteriors on certain galaxy properties.
The mock observations are constructed from \lgal~and include photometry and
spectra. 
In the mock spectra, we model the fiber aperture effect --- \emph{i.e.} spectra
only include light from a galaxy collected within its fiber diameter --- by
scaling the SED flux (Section~\ref{sec:spec}).
In our SED modeling, we account for this fiber aperture effect using a
normalization factor, $f_{\rm fiber}$ (Section~\ref{sec:sps}). 
Hence, our mock observations and SED modeling have a consistent treatment of
the fiber aperture effect. 
In observations, however, aperture effects can be wavelength
dependent~\citep{gerssen2012, richards2016} and if the dependence is strong,
an overall $f_{\rm fiber}$ factor would not be sufficient.
We examine the wavelength dependence for BGS by comparing the ratio of the
fiber aperture flux over total flux, $f^{\rm fiber}_X/f_X$, in $g$, $r$, and
$z$ bands of BGS targets from LS.
We find find no significant difference in the flux ratios of the different
bands, which suggests that the fiber aperture effect does not have a strong
wavelength dependence for BGS galaxies. 

Flux calibration performed by the DESI spectral pipeline can also induce
wavelength dependent residuals. 
DESI spectra are measured using three-arm spectrographs that split the spectra
into three $b$, $r$, and $z$ channels with overlapping wavelength ranges: 
$3600 - 5930$, $5660 - 7720$, and $7470 - 9800 \AA$.  
After flat fielding and sky subtraction, flux calibration is performed on each
channel of the spectra by matching physical stellar models to spectra of
spectrohotometric standard stars observed in the same exposure
(Guy~\etal~in prep.). 
Since the calibration is performed for each channel separately, imperfections
can imprint a wavelength dependent residual. 
In a subsequent paper, Ramos \etal~(in prep.), we examine the fiber aperture
effect and wavelength dependent imprints on DESI spectra using BGS
spectra from the DESI Survey Validation data and observations from the Mapping
Nearby Galaxies at APO (MaNGA) survey. 
Using galaxy properties derived using the {\sc PROVABGS} pipeline for spectra
from integrated field unit MaNGA observations, we will present aperture
corrections that can be applied on derived BGS galaxy properties. 
We also note that the {\sc PROVABGS} SED modeling pipeline already includes flux
calibration models beyond a single $f_{\rm fiber}$ and can easily be extended
to include more sophisticated models~\citep[\emph{e.g.} Chebyschev
polynomial;][]{carnall2019a, tacchella2021}. 

%\todo{paragraph on how we handle theoretical assumptions --- isochrones and stellar libraries, summary of the appendix}
\subsection{Stellar Model Choices}
In both our {\sc PROVABGS} SED model and mock observations, we use the MIST
isochrones, the combined MILES+BaSeL spectral library, and the
\cite{chabrier2003} IMF.
With the same set of choices, our analysis does not consider how different
choices for stellar evolution or IMF can affect the inferred galaxy properties. 
Yet, it is well-established that there are major uncertainties in each of these
choices~\citep{conroy2009, conroy2013}.
For instance, recent observational works suggest that there may be significant
variations in IMF~\citep[\emph{e.g.}][]{treu2010, vandokkum2010, rosani2018,
sonnenfeld2019}. 
Different SPS model choices can also significantly impact the derived galaxy
propeties~\citep[\emph{e.g.}][]{ge2019}.
We reserve a detailed examination of this effect for future work. 
In the meantime, for the {\sc PROVABGS} catalog we will release multiple
catalogs each with different sets of choices for isochrone, spectral library,
and IMF.

\subsection{Advantages of PROVABGS}
We demonstrate with the mock challenge that we can derive accurate and precise
constraints on specific galaxy properties using the {\sc PROVABGS} SED modeling.
The {\sc PROVABGS} catalog will have a number of key advantages over other
value-added galaxy catalogs. 
First, {\sc PROVABGS} will provide full Bayesian posteriors on galaxy
properties instead of ``best-fit'' point estimates from maximizing the
likelihood. 
Posterior distributions are essential for accurately estimating uncertainties
on galaxy properties.  
These uncertainties are significant, especially for properties such as $\zmw$
(Figure~\ref{fig:prop_inf}). 
Ignoring them dramatically overestimates the statistical precision of the
derived galaxy properties and can significantly bias any galaxy study.
%We also note that the maximum-entropy method, mentioned earlier, to correct for the effect of priors on derived galaxy properties requires full posterior distributions.

Furthermore, the {\sc PROVABGS} posteriors will be derived from MCMC sampling
rather than grid-based methods often used in the
past~\citep[\emph{e.g.}][]{dacunha2008, moustakas2013, boquien2019}.
As a result, they can accurately estimate posterior distributions with
significant parameter degeneracies or multiple modes (peaks). 
For instance, in the posterior of Figure~\ref{fig:posterior} we find
degeneracies between $f_{\rm burst}$ and $\{\beta_1, \beta_2, \beta_3,
\beta_4\}$ and between $\{\gamma_1, \gamma_2\}$ and $\{\beta_1, \beta_2,
\beta_3, \beta_4\}$. 
The posterior is also multi-modal. 
Accurate estimates of the full posterior distribution are especially important,
as they enable the maximum-entropy method, mentioned earlier, 
to correct for the significant impact of priors on derived galaxy properties.
Grid-based methods also scale exponentially with the number of SPS parameters
so they quickly become infeasible as the dimensionality of SPS models increase. 
MCMC, on the other hand, scales approximately linearly with the number of
parameters. 

\begin{figure}
\begin{center}
\includegraphics[width=0.7\textwidth]{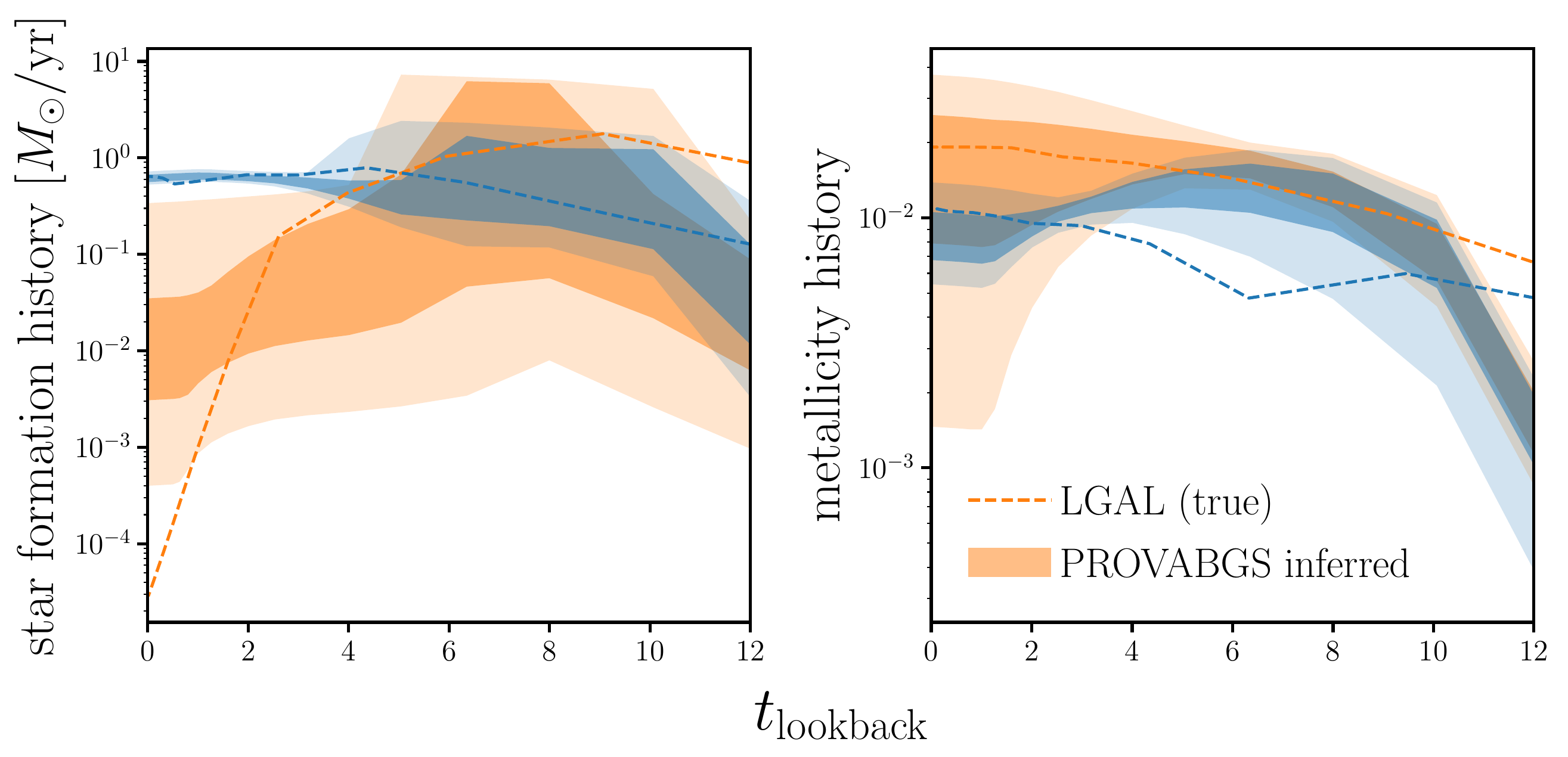}
    \caption{
        With the {\sc PROVABGS} SPS model, we can infer posteriors on the full
        star formation and metallicity histories. 
        We present the inferred SFH and ZH for an arbitrarily chosen
        star-forming (blue) and quiescent galaxy (orange).
        The shaded region represent the 64 and 95\% confidence intervals of the
        SFH and ZH posteriors. 
        For comparison, we include the true SFH and ZH (dashed). 
        The inferred SFH and ZH show good agreement with the true values;
        however, similar to the inferred $\avgsfr$ and $\zmw$, the SFH and ZH
        are significantly impacted by priors imposed by the SPS model. 
    } \label{fig:sfh_demo}
\end{center}
\end{figure}
%\todo{Beyond the galaxy properties we discuss in Section~\ref{sec:results}, we can also derive SFH and ZH}

In this work, we primarily focus on the following physical properties of
galaxies: $\log M_*$, $\log\avgsfr$, $\log\zmw$, $\tage$, and $\tauism$. 
The {\sc PROVABGS} SPS model, however, can constrain galaxy properties beyond
these properties. 
%The SPS model employs nonparametric SFH and ZH prescriptions based on NMF bases and the model parameters include coefficients for these bases. 
Posteriors on the SPS model parameters can, thus, be used to derive constraints
on the SFH and ZH. 
In Figure~\ref{fig:sfh_demo}, we present the inferred SFH and ZH of two
simulated galaxies from our \lgal~sample: a star-forming (blue) and a quiescent
galaxy (orange). 
We mark the 68 and 95\% confidence intervals in the shaded regions. 
For comparison, we include the true SFH and ZH from \lgal~(dashed).  
The inferred SFH and ZH is able to generally recover the true histories. 
We emphasize that current SPS models typically assume constant ZHs that does
not vary over time~\citep{carnall2019a, leja2019}. 
Hence inferring ZH over time is a key advantage of the {\sc PROVABGS} SPS
model. 
Similar to the inferred $\avgsfr$ and $\zmw$, the SFH and ZH constraints are
also impacted by the priors imposed by our SPS model
(Appendix~\ref{sec:model_priors}, Figure~\ref{fig:sfh_prior}).

Another key advantage of {\sc PROVABGS} is that it will infer galaxy
properties from joint SED modeling of photometry \emph{and spectra}. 
Our results illustrate the advantages of including spectra in SED modeling. 
Galaxy spectra provide substantial statistical power for constraining 
galaxy properties. 
In addition to tightening constraints overall, their statistical power is
essential for mitigating the effect of the model priors. 
For instance, including spectra in the SED modeling significantly reduces the
bias of our $\zmw$ and $\tage$ constraints (Figure~\ref{fig:etas}). 
It also reduces the lower bound on the inferred $\avgsfr$. 
In fact, without spectra, we are dominated by priors on $\avgsfr$ and cannot
robustly infer galaxy properties of quiescent galaxies with $\log\avgsfr < 0$
dex.
%We emphasize that all of these benefits come from spectra with the SNR of BGS, which is observing during bright time. 

\subsection{Applications of PROVABGS}
{\sc PROVABGS} will be a value-added galaxy catalog with unprecedented
statistical power. 
With physical galaxy properties of over $10$ million DESI BGS galaxies, 
{\sc PROVABGS} will provide a transformational galaxy sample to extend
previous statistical galaxy studies. 
For example, we will be able to make the most precise measurement of the
stellar mass function~\citep[SMF]{li2009, moustakas2013}, star-forming
sequence~\citep{noeske2007, curtis-lake2021}, mass-metallicity
relation~\citep{tremonti2004}, or any other summary statistic of galaxy
populations. 
{\sc PROVABGS} will also include large sample of dwarf galaxies thanks to the
faint apparent magnitude limit of BGS. 
Dwarf galaxies are dark matter dominated and, thus, probe the physics of
dark matter; they are also sensitive to star formation feedback and can help
distinguish different aspects of galaxy formation~\citep{mao2021}. 
Galaxy studies examining the galaxy-halo connection can also be extended to
exploit the additional statistical power of {\sc
PROVABGS}~\citep[\emph{e.g.}][]{tinker2011, wetzel2013, zu2015, hahn2017,
hahn2019b}. 
With detailed galaxy properties, {\sc PROVABGS} will also enable
multiple-tracer galaxy clustering analyses that can circumvent cosmic variance
in inferring cosmological parameters~\citep{seljak2009, mcdonald2009,
wang2020}.
Analyses exploiting new forward modeling approaches, such as \cite{hahn2021},
will also greatly benefit from the statistical power of {\sc PROVABGS}.
%\ch{Text above will be updated with more science applications based on any feedback.}

In addition to the applications above, {\sc PROVABGS} will also unlock
applications that can exploit the full posteriors of the probabilistic catalog.
In this work, we utilized the posteriors in order to quantify accuracy and
precision of galaxy population constraints using population inference with a
hierarchical Bayesian approach. 
This is only the \emph{simplest} illustration of such an approach. 
Another application is to use posteriors on $M_*$, $p(M_* \given {\bfi X}_i)$,
to measure $p(M_* \given \{ {\bfi X_i} \})$ --- the \emph{probabilistic} SMF.
With full posteriors, we can probe even the lowest signal-to-noise regime
accurately so the SMF will be reliable at the lowest mass end, down to
${\sim}10^{7} M_\odot$ (Figure~\ref{fig:bgs_mstar}). 
This will constrain the SMF of dwarf galaxies and have important
implications for both galaxy evolution and cosmology. 

%With the {\sc PROVABGS} posteriors we can infer fully probabilistic galaxy
%population statistics, which will allow us to robustly probe even the lowest
%signal-to-noise regime.
%A probabilistic SMF of BGS, for example, will provide accurate constraints on
%the low mass end down to ${\sim}10^{7} M_\odot$ (Figure~\ref{fig:bgs_mstar}),
%which has important implications for both galaxy evolution and cosmology. 
Probabilistic analyses can extend to higher dimensions. 
Joint posteriors on $M_*$ and SFR, $p(M_*, {\rm SFR} \given {\bfi X}_i)$
can be used to measure the probabilistic star formation sequence. 
Since the posteriors reliably estimate the uncertainties and parameter
degeneracies, we will more accurately infer the intrinsic width of the
SFS, which encodes information about star formation and stellar and AGN
feedback in galaxies~\citep{davies2021}. 
We can even extend the approach to infer the distribution of \emph{all}
galaxy properties given observations, $p(\theta | \{X_i\})$, which would
exploit the \emph{full} statistical power of observations and reveal new
trends among galaxy properties. 
This is only possible with population inference using the posterior
distributions of every galaxy.

%We can also use population inference to robustly derive galaxy property distributions of galaxy subpopulations --- \emph{without stacking observations}.
Population inference also allows us to avoid stacking observations. 
Stacking makes the strong assumption that galaxies that are grouped together in
some \emph{e.g.} color-space are from a subpopulation with the same properties. 
This assumption fails if, for instance, there are contaminants or multiple
disparate galaxy subpopulations that are degenerate in color-space and
therefore are included in the stack. 
With all of the applications listed above, {\sc PROVABGS} will enable us to
fully extract the statistical power of >10 million BGS galaxies.

% --- summary ---  
\section{Summary}
Over the next five years, DESI will measure spectra for ${>}30$ million
galaxies, each with optical photometry from the Legacy Surveys. 
BGS, which will extend out to $z\sim0.6$, will provide a $r < 19.5$
magnitude-limited sample of ${\sim}10$ million galaxies spanning a wide range
of galaxy properties with high completeness.
It will also include a sample of ${\sim}5$ million fainter galaxies down to $r
< 20.175$ selected based on a fiber magnitude and color. 
This upcoming dataset offers a unique opportunity to leverage its statistical
power for galaxy evolution and maximize its scientific impact. 
Accurate galaxy properties for such a galaxy sample, for instance,  would
enable us to measure population statistics and empirical relations of galaxies
with unprecedented precision. 
It would also enable more complete and precise comparisons between observations
and galaxy formation models, which will shed light into the physical processes
of galaxy evolution.
To exploit this opportunity, we will construct the PRObabilistic Value-Added
Bright Galaxy Survey (PROVABGS) catalog, where we will apply state-of-the-art
Bayesian SED modeling to jointly analyze DESI photometry and spectroscopy. 
PROVABGS will provide full posterior distributions of galaxy properties, such as
stellar mass ($M_*$), star formation rate (SFR), stellar metallicity 
($\zmw$), and stellar age ($\tage$), for all ${>}10$ million BGS galaxies.

In this work, we present and validate the SED model, Bayesian inference
framework, and other methodology that will be used to construct
PROVABGS\footnote{publicly available at
\href{https://github.com/changhoonhahn/provabgs/}{https://github.com/changhoonhahn/provabgs/}}.
We use 2,123 galaxies in the {\sc L-Galaxies} semi-analytic model to construct
realistic synthetic DESI spectra and photometry.  
We build SEDs using SPS based on the star formation and chemical enrichment
histories of the simulated galaxies.
Then, we simulate the SEDs using the forward modeling pipeline used in the BGS
survey design.  
Afterwards, we apply the PROVABGS SED modeling on the mock DESI observations to
derive posteriors on $M_*$, $\avgsfr$, $\zmw$, and $\tage$. 
From the posteriors and the population inference we conduct to quantify
accuracy and precision, we find: 
\begin{itemize}
    \item Overall, we derive posteriors of galaxy properties that are in good
        agreement with the true properties of the simulated galaxies. 
        Furthermore, with posteriors rather than point estimates we accurately
        estimate the uncertainties on the galaxy properties. 
        We infer posteriors with the following levels of precision: 
        $\sigma_{\log M_*}\sim0.1$ dex, $\sigma_{\log\avgsfr}\sim0.1$ dex, 
        $\sigma_{\log\zmw}\sim0.15$ dex, and $\sigma_{\tage}\sim0.5$ Gyr. 
        Our results also demonstrate that we successfully marginalize over the
        effect of dust and other nuisance parameters. 
    \item Like any SED model, the PROVABGS SED model imposes significantly
        non-uniform priors on galaxy properties. 
        We find that these priors impose a lower bound on $\avgsfr$ of 
        $\avgsfr > 10^{-1}M_\odot/{\rm yr}$. 
        It also biases $\zmw$ by ${\sim}0.3$ dex for observations with low
        spectral signal-to-noise and imposes an upper bound of $\tage < 8$ Gyr. 
        We characterize the priors in detail so that constraints on galaxy
        properties can be interpreted in future studies that use PROVABGS.
    \item We compare the posteriors derived from DESI spectrophotometry to
        those derived from photometry alone. 
        Including DESI spectra substantially improves the constraints on galaxy
        properties. 
        Moreover, jointly analyzing spectra is {\em essential} for mitigating
        the impact of the SED model priors. 
        For example, with photometry alone, the priors impose a more
        restrictive $\avgsfr > 1 M_\odot/{\rm yr}$ lower bound and bias $\zmw$
        ${\sim}0.5$ dex.
\end{itemize}

We demonstrate with our mock challenge that we will derive accurate and precise
constraints on specific galaxy properties in PROVABGS. 
Beyond $M_*$, $\avgsfr$, $\zmw$, and $\tage$, which we focus on in this work, 
PROVABGS will also constrain star formation and metallicity histories. 
With galaxy properties of over ${>}10$ million BGS galaxies, current galaxy
studies will be able to use the PROVABGS catalog to exploit the statistical
power of BGS for the most precise measurements of various galaxy relations.
Since the BGS samples span a wide range of galaxies, PROVABGS will
also enable galaxy studies to investigate less explored regimes, such as dwarf
galaxy populations. 

Furthermore, PROVABGS will be a fully probabilistic catalog.
With posteriors for all the galaxy properties, we can conduct more rigorous
statistical analyses using new techinques such as population inference and
hierarchical Bayesian modeling.
We demonstrate one such approach in this work by using population inference to
estimate the overall accuracy and precision of our galaxy property constraints. 
These methods will not only improve the accuracy of our analyses but they will
also allow us to fully exploit the statistical power of DESI observations. 

Despite the overall success of the PROVABGS methodologies that we demonstrate,
there are some limitations. 
For instance, we only consider a simple model for the effect of the DESI fiber
aperture and flux calibration. 
A more detailed investigation will be presented in Ramos~\etal~(in prep.). 
We also do not consider varying the isochrones, stellar library, or IMF. 
Instead, we will release multiple versions of PROVABGS with different sets of
assumptions. 
Lastly, we find that the most significant limitation to deriving accurate
galaxy properties comes from the prior imposed by the SED model. 
We will address this limitation and present a method to impose uniform priors
on galaxy properties in Hahn~(in prep.). 

DESI has started its main 5 year operation. 
Already, as part of survey validation, DESI has collected over 400,000 spectra
of BGS galaxies that will be released in the Survey Validation Data Assembly
(SVDA). 
The SVDA release will also be accompanied by papers describing the data
reduction pipeline, redshift fitting algorithm, fiber assignment, survey
operation and simulations, visual inspection, and target selection for the
various tracers. 
Finally, using BGS observations in the SVDA, we will construct and release the
PROVABGS-SV catalog and present the probabilistic stellar mass function
measured from it in the subsequent paper.

The entire PROVABGS SED modeling pipeline, including the neural emulators and
Bayesian inference framework, is publicly available at:
\url{https://github.com/changhoonhahn/provabgs/}.
All of the software and scripts used in our analysis are publicly available at:
\url{https://github.com/changhoonhahn/gqp_mc}.
The accompanying data used in this work, including the mock DESI observations
and posteriors derived from PROVABGS, is available at:
\url{https://doi.org/10.5281/zenodo.5910635}.

\section*{Acknowledgements}
It's a pleasure to thank
    Justin Alsing, 
    Adam Carnall, 
    Charlie Conroy, 
    Kartheik Iyer, 
    Stephanie Juneau, 
    Joel Leja, 
    Jenny Greene, 
    Peter Melchior,
    Michael A. Strauss
for valuable discussions and comments. 
The authors would also like to thank Song Huang for valuable feedback and
comments during the DESI internal review. 
This material is based upon work supported by the U.S. Department of Energy,
Office of Science, Office of High Energy Physics, under contract No.
DE-AC02-05CH11231.  This project used resources of the National Energy Research
Scientific Computing Center, a DOE Office of Science User Facility supported by
the Office of Science of the U.S.  Department of Energy under Contract No.
DE-AC02-05CH11231. 
CH is supported by the AI Accelerator program of the Schmidt Futures Foundation.
MS is supported by the European Union's  Horizon 2020 research and innovation
programme under the Maria Sk\l{}odowska-Curie (grant agreement No 754510), the
National Science Centre of Poland (grant UMO-2016/23/N/ST9/02963) and by the
Spanish Ministry of Science and Innovation through Juan de la Cierva-formacion
program (reference FJC2018-038792-I).
MM acknowledges support from the Ramon y Cajal fellowship (RYC2019-027670-I).

This research is supported by the Director, Office of Science, Office of High
Energy Physics of the U.S. Department of Energy under Contract No.
DE–AC02–05CH11231, and by the National Energy Research Scientific Computing
Center, a DOE Office of Science User Facility under the same contract;
additional support for DESI is provided by the U.S. National Science
Foundation, Division of Astronomical Sciences under Contract No. AST-0950945 to
the NSF’s National Optical-Infrared Astronomy Research Laboratory; the Science
and Technologies Facilities Council of the United Kingdom; the Gordon and Betty
Moore Foundation; the Heising-Simons Foundation; the French Alternative
Energies and Atomic Energy Commission (CEA); the National Council of Science
and Technology of Mexico; the Ministry of Economy of Spain, and by the DESI
Member Institutions.

The authors are honored to be permitted to conduct scientific research on
Iolkam Du’ag (Kitt Peak), a mountain with particular significance to the Tohono
O’odham Nation.

\appendix
\section{Non-negative Matrix Factorization Bases} \label{sec:nmf}
The basis vectors for the star-formation and metallicity histories are computed
using non-negative matrix factorisation (NMF) on a set of star formation and
metallicity histories in the Illustris 
simulation~\citep{vogelsberger2014, genel2014, nelson2015}.
Unlike PCA, NMF lends itself well to this task as it gives positive vectors,
which can each be straightforwardly interpreted physically as representing the
SFH of a composite stellar population. 
In the case of the ZHs, the advantage of NMF over PCA is less clear, but we
maintain the NMF scheme for simplicity. 

The SFHs and ZHs are computed from all stellar particles bound to subhalos
that host a galaxy with $M_* > 10^9 M_\odot$ at $z=0$, giving a sample of just
over 29,000 Illustris galaxies. 
For the SFHs, we take the distribution of stellar ages in 400 bins,
logarithmically distributed between 8.6 Myrs and 13.65 Gyrs, and compute the
stellar mass formed in each bin. For the ZHs, we take the mass-weighted
metallicity in each of the bins. 
Next, the vectors for the SFHs and ZHs are normalized independently ---
\emph{i.e.} we do not keep information of which ZH corresponds to each SFH.
Therefore we do not impose the mass-metallicity relation of the simulation onto
our basis vectors (see \citealt{thorne2021} for a parameterization that links
SFH with ZH throught he mass-metallicity relation). 
We take each set of simulated SFHs and ZHs as a reasonable representation of
possible SFHs and ZHs in the Universe. 
Prior to decomposition, each individual vector is smoothed on a scale of 400
Myr, which removes any information on smaller timescales. 
We decompose the set of SFHs into 4 independent components, and the set of ZHs
into 2 independent components. 
The resulting components are shown in the main text (Figure~\ref{fig:nmf}). 

\begin{figure}
\begin{center}
\includegraphics[width=0.85\textwidth]{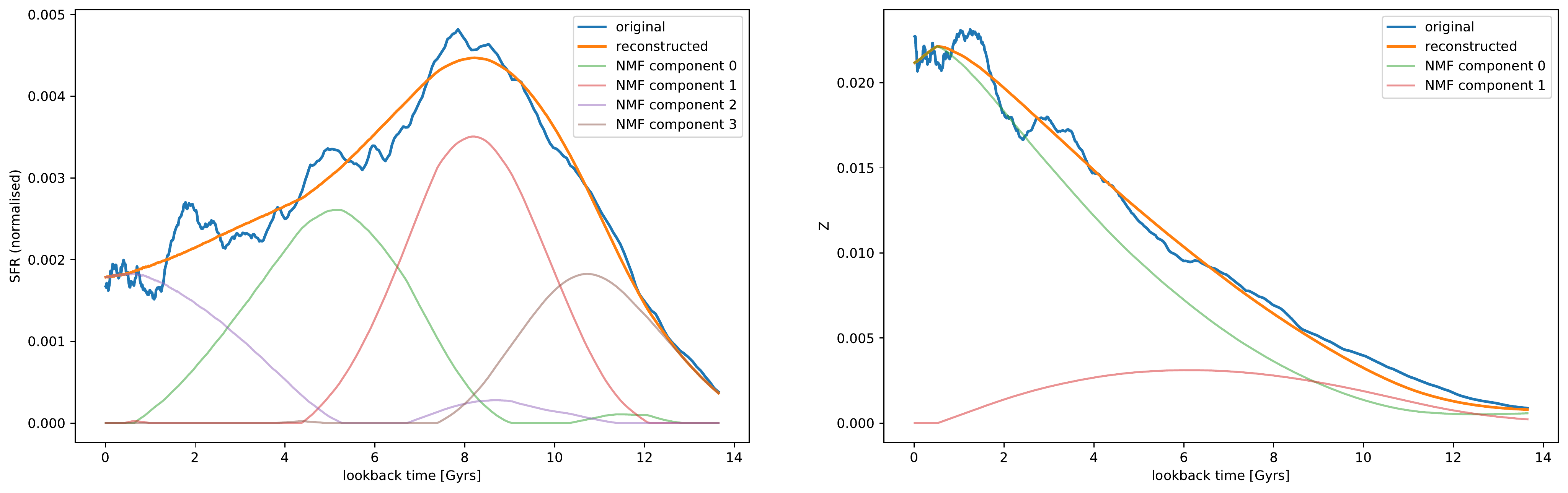}
\includegraphics[width=0.85\textwidth]{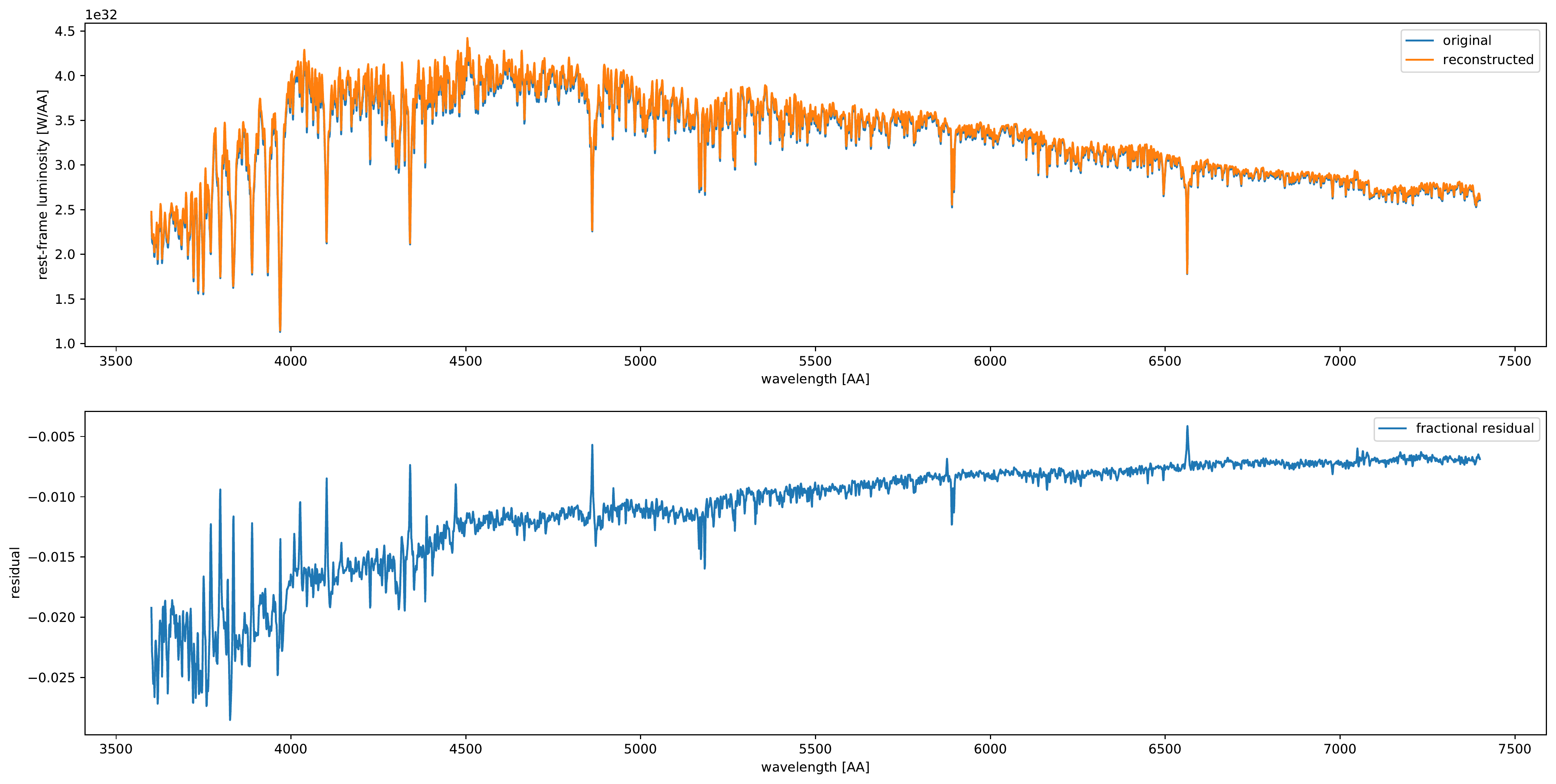}
    \caption{
    The original and NMF-reconstructed SFHs (top left) and ZHs (top right) of a
    galaxy in the Illustris simulation. 
    The original SFH is shown after smoothing on a scale of 400 Myr. 
    We mark the contributions of each of the NMF components in the faded
    colored lines. 
    The middle and bottom panels compare the spectra obtained from integrating
    the original and reconstructed SFH and ZH. 
    In this case, the NMF basis offers a good reconstruction of the SFH and ZH,
    which results in a small residuals in the corresponding spectra.
    }\label{fig:nmf0}
\end{center}
\end{figure}

\begin{figure}
\begin{center}
\includegraphics[width=0.85\textwidth]{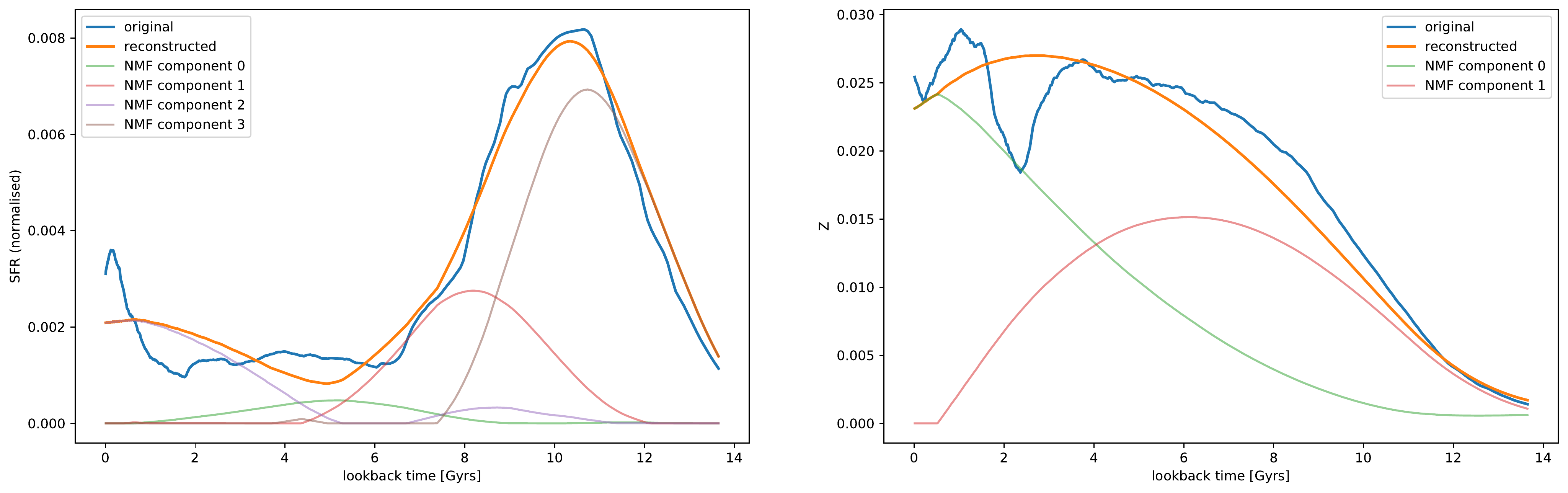}
\includegraphics[width=0.85\textwidth]{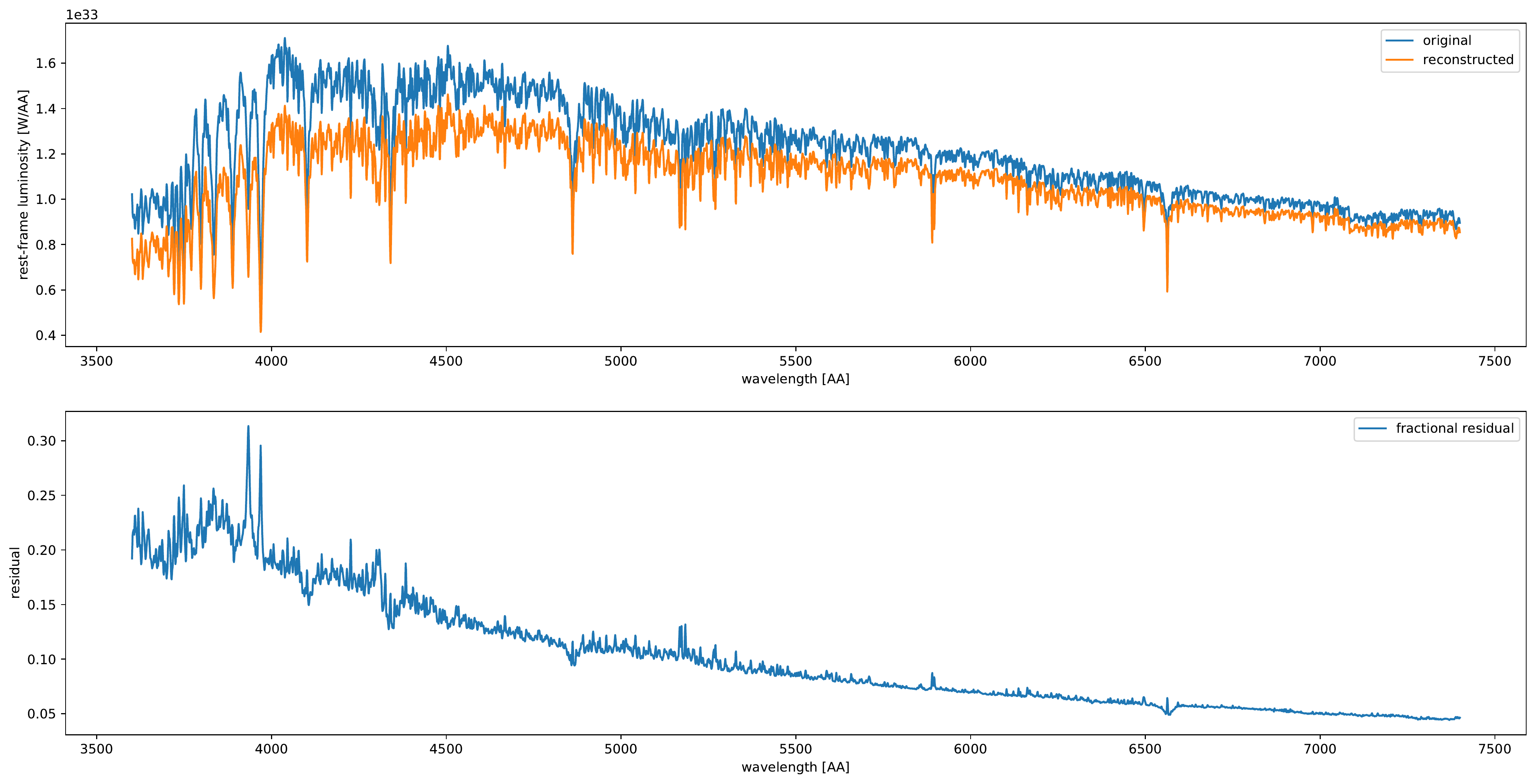}
    \caption{
    Same as Figure~\ref{fig:nmf0} but for another Illustris galaxy. 
    In this case, the NMF basis fails to reproduce a burst of star formation at
    recent times, leading directly to an underestimation of the luminosity,
    especially towards the bluer wavelengths.
    }\label{fig:nmf1}
\end{center}
\end{figure}

\begin{figure}
\begin{center}
\includegraphics[width=0.85\textwidth]{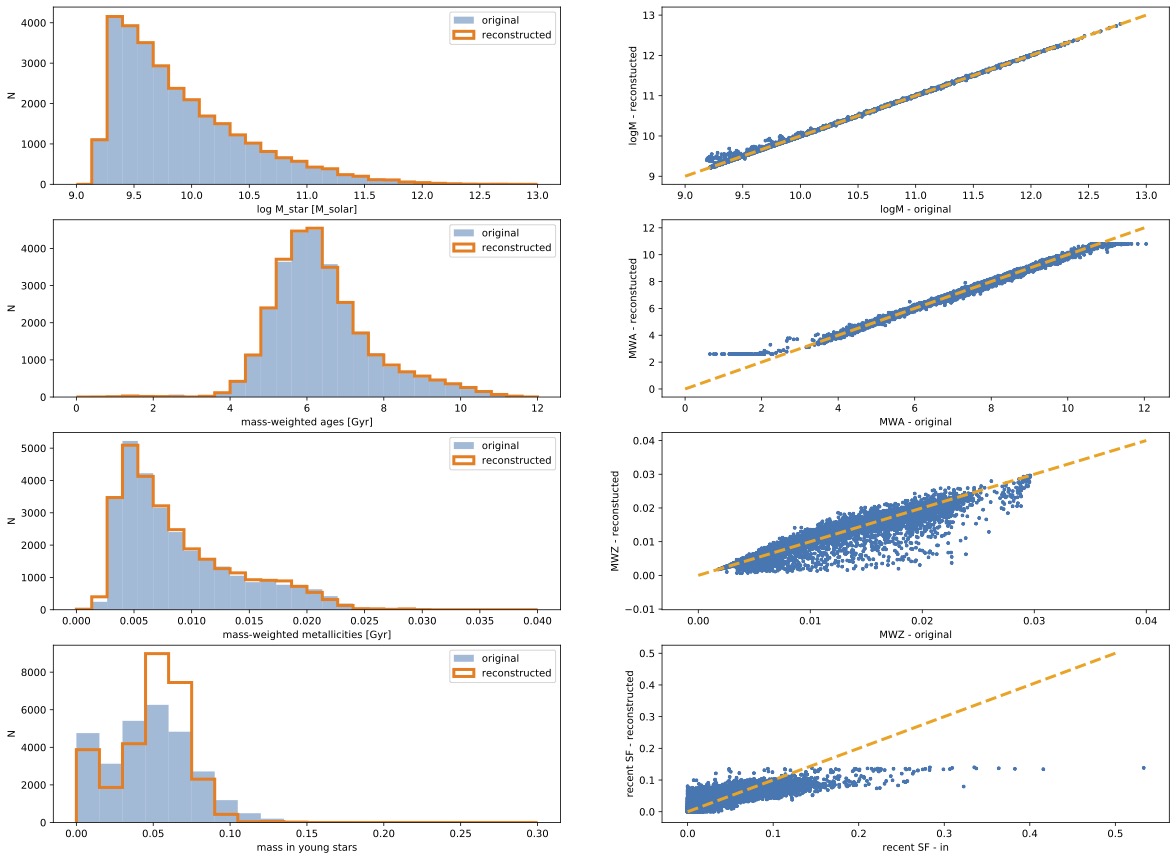}
    \caption{
    Comparison of original versus NMF reconstructed total stellar mass formed,
    mass-weighted age, mass-weighted metallicity, and mass formed in the last
    200 Myr (from top to bottom). 
    On the left column we present the distribution of each quantity, and on the
    right column we show direct comparisons in scatter plots. 
    The orange dashed lines shows the one-to-one line.
    Total stellar mass is well recovered, as expected given its lack of
    sensitivity to smaller bursts. 
    Mass-weighted ages are poorly recovered at young and old ages, as a direct
    consequence of the lack of resolution of our basis.
    The mass-weighted metallicity is well recovered on the mean, though with a
    large scatter. 
    The mass formed in young stars is again affected by the lack of resolution
    of our basis. 
    In our SPS model, we include a stochastic burst component to account for
    this limitation (Section~\ref{sec:sps}). 
    }\label{fig:nmf2}
\end{center}
\end{figure}

Figures~\ref{fig:nmf0} and~\ref{fig:nmf1} show two examples of the NMF direct
reconstruction on two galaxies. 
The two galaxies are chosen as examples of a `fair' and a `poor' reconstruction.
In all cases the reconstructions can be improved by increasing the number of
components, and doing so effectively improves our ability to model shorter
timescale features in the SFH and ZHs.
In this work, we instead include a stochastic burst component in the SFH
(Section~\ref{sec:sps}). 

In Figure~\ref{fig:nmf2}, we present how NMF reconstruction projects onto
certain derived properties: total stellar mass formed, mass-weighted age,
mass-weighted metallicity and mass in young stars (mass formed in the last 200
Myr).
Besides the total stellar mass, the other derived properties are impacted by
the lack of short timescale features. 
Our stochastic burst component directly addresses this limitiation. 
Therefore, the NMF basis can be seen as a reasonable and minimal set to recover
the broad shape of the star-formation and metallicity histories, which is
complemented in our SPS model by the stochastic burst component.

\section{SPS Model Priors} \label{sec:model_priors}
SED models impose undesireable non-uniform priors on galaxy properties that
significantly impact their ability to infer unbiased galaxy properties such as
$\avgsfr$, $\zmw$, and $\tage$. 
For the PROVABGS SED modeling we present in this work, model imposed priors
place a lower bound on $\avgsfr$, bias $\zmw$ for observations with low
spectral SNR, and place an upper bound of $\tage < 8$ Gyr.
Given their significant impact, we quantify and characterize the model imposed
prior in further detail below. 

% explanation of how the parameters are derived 
Model imposed priors are a consequence of the fact that many of the galaxy
properties of interest are not explicit parameters of the SPS model. 
Out of the properties we focus on in this work, only $M_*$ is a parameter in
our SPS model. 
$M_*$ determines the overall amplitude of the SED. 
Meanwhile, $\avgsfr$ is derived from integrating the SFH over 
$t_{\rm age} - 1\,{\rm Gyr} < t < t_{\rm age}$.
The SFH is itself a derived quantity from the SED model parameters $\{\beta_1,
\beta_2, \beta_3, \beta_4 \}$, $f_{\rm burst}$, $t_{\rm burst}$, and $M_*$
(Eq.~\ref{eq:nmf} and~\ref{eq:sfh}).
$\zmw$ is an even more complicatedly derived quantity that involves integrating
the product of the SFH and ZH, and hence depends on $\{\beta_1, \beta_2,
\beta_3, \beta_4 \}$, $f_{\rm burst}$, $t_{\rm burst}$, and $\{\gamma_1,
\gamma_2\}$ (Eq.~\ref{eq:prop_eqs}).
$\tage$ is similarly derived by integrating the SFH by age. 
All of these derived properties are further impacted by the fixed log-spaced
$\tlb$ binning since the integrals are evaluated discretely
(Section~\ref{sec:sps}). 

\begin{figure}
\begin{center}
\includegraphics[width=0.75\textwidth]{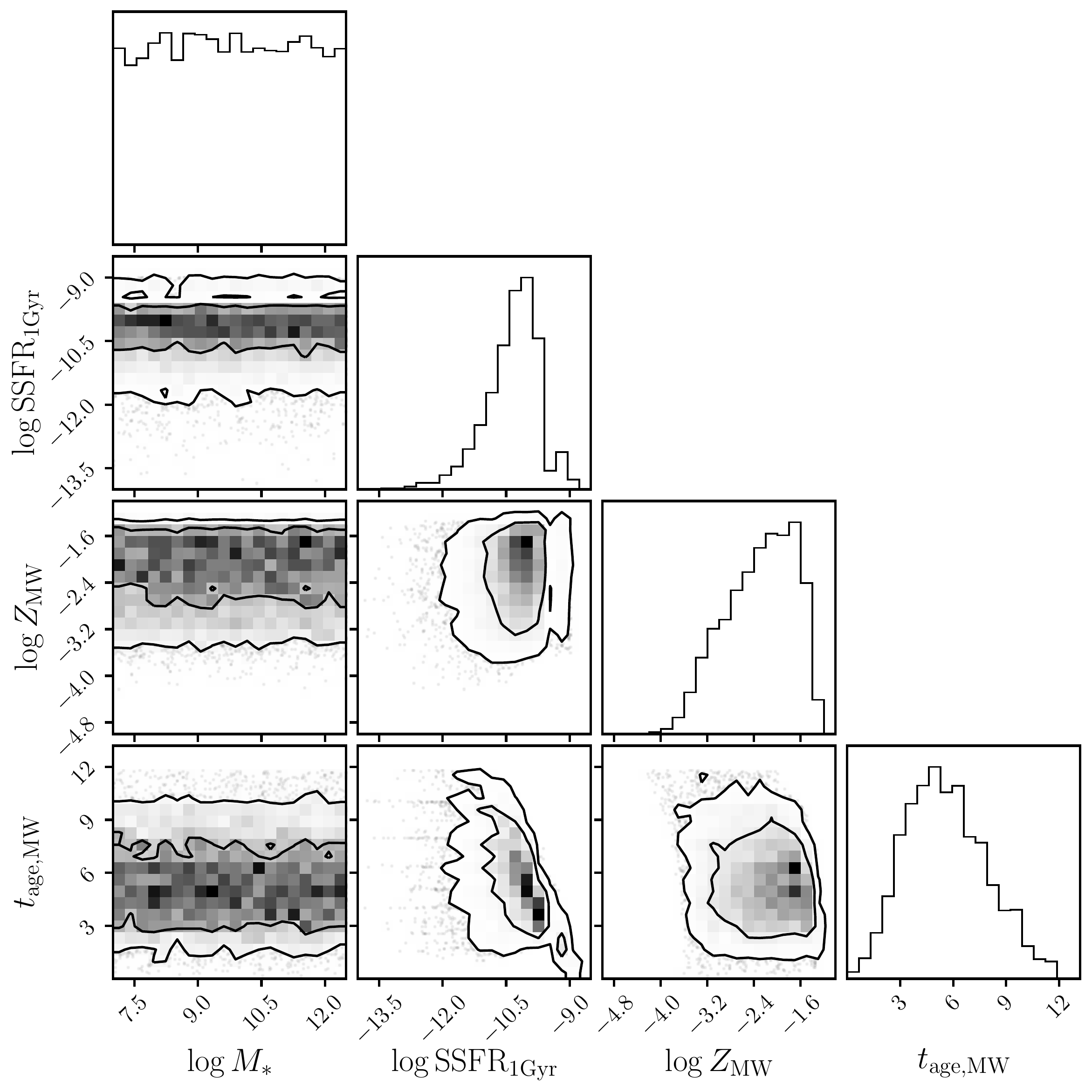}
    \caption{
    Priors imposed by our SPS model on galaxy properties $\log M_*$, $\log
    \overline{\rm SSFR}_{\rm 1 Gyr}$, $\log Z_{\rm MW}$, and $\tage$ at
    $z=0.1$. 
    Out of the galaxy properites, only $\log M_*$ is a parameter in our SPS
    model.
    The others are derived from the SPS model parameters
    (Eq.~\ref{eq:prop_eqs}). 
    Hence, even when we impose uninformative priors on the model parameters as
    in Table~\ref{tab:params}, we do not impose uniform priors on the galaxy
    properties. 
    In fact, for $\avgssfr$, $\zmw$, and $\tage$, our SPS model imposes
    significantly skewed distributions that explain the biases and bounds we
    find in the galaxy property posteriors.  
    All SPS models impose undesireable priors on galaxy properties. 
    By characterizing the prior above, we provide a way to interpret the
    posteriors on galaxy properties for PROVABGS and disentangle the effect of
    the prior. 
    }\label{fig:model_prior}
\end{center}
\end{figure}
% present the figure and describe in detail how we derive the figure. 
We illustrate and quantify the model imposed priors on the galaxy
properties in Figure~\ref{fig:model_prior}. 
We present the probability distribution of the priors on $\log M_*$,
$\log\avgssfr$, $\log\zmw$, and $\tage$ for galaxies at $z=0.1$.
The distribution is derived by first sampling SPS parameters from prior
specified in Table~\ref{tab:params}, $\theta'\sim p(\theta)$.
Then for each $\theta'$, the galaxy properties are derived using
Eq.~\ref{eq:prop_eqs}. 
We present $\log\avgssfr = \log(\avgsfr/M_*)$ instead of $\log\avgsfr$ to
remove the correlation with $M_*$.  
The contours mark the 68 and 95\% of the distribution. 
We note that the prior distribution depends on redshift since it determines
$t_{\rm age}$.
The dependence is relatively small over the BGS $z$ range so we only show
$z=0.1$ for simplicity. 

% characterize the prior distribution 
We confirm that the prior on $\log M_*$ is uniform as we specify in
Table~\ref{tab:params}. 
For the other parameters, however, the model imposed prior is \emph{not}
uniform. 
For $\avgssfr$, the prior spans $-13.5 < \log \avgssfr < -9$; however, it skews
toward the primary peak at $\log\avgssfr = -10.4$. 
The secondary peak near $-9$ dex is a consequence of the starburst component that
we include in the SFH. 
By definition $\avgssfr$ cannot exceed $10^{-9}\,yr^{-1}$. 
For $\log\zmw$ and $\tage$, the priors are also skewed distributions that peak
near -1.6 dex and 6 Gyr, respectively.
Furthermore, for $\tage$, the prior reveals the imprint of the log-spaced 
$\tlb$ bins (see $\tage$ versus $\log\avgssfr$ panel).
As we discuss in the main text, the shape of the model imposed priors on
$\avgssfr$, $\zmw$, and $\tage$ explains the limitations of the posteriors we
derive from our SED modeling.

\begin{figure}
\begin{center}
\includegraphics[width=0.4\textwidth]{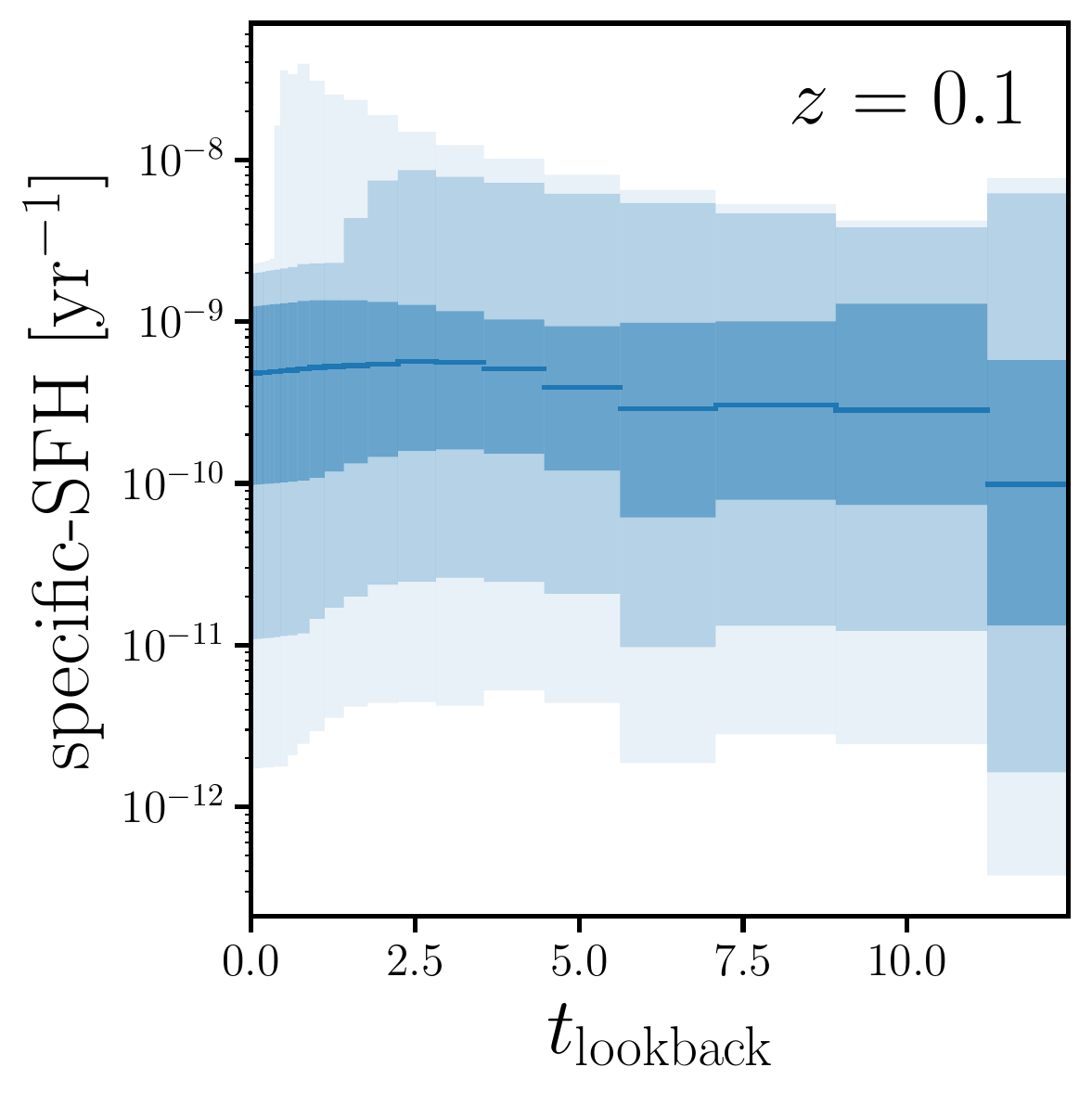}
    \caption{
        Priors imposed by our SPS model on the specific-SFH (sSFH) for galaxies
        at $z=0.1$. 
        We represent 68, 95, 99.7\% of the sSFH distribution with the shaded
        regions (dark to light). 
        Our SPS model imposes a prior on sSFH that is asymmetric and peaked at 
        ${\sim}5\times10^{-10}{\rm yr}^{-1}$. 
        For observations where the likelihood distribution is diffuse
        (\emph{e.g.} low SNR), the inferred SFH will be significantly skewed
        towards the peak of the distribution. 
        Overall, this prior will cause the inferred SFHs to be flatter over
        $\tlb$ and skew towards the peak as we see in the comparison between
        the inferred and true SFHs in Figure~\ref{fig:sfh_demo}. 
        Any analysis of SFHs based on SED modeling must take account the effect
        of model imposed priors. 
    }\label{fig:sfh_prior}
\end{center}
\end{figure}
In addition to the galaxy properties above, we also characterize the model
imposed prior on specific-SFH (sSFH) in Figure~\ref{fig:sfh_prior}. 
The sSFH is the SFH normalized by total stellar mass. 
The shaded regions represent 68, 95, 99.7\% of the SFH distribution (dark to
light). 
We show the prior for galaxies at $z=0.1$. 
Throughout the $\tlb$ range, the sSFH prior is asymmetric and peaks at
${\sim}5\times10^{-10}{\rm yr}^{-1}$. 
Since this prior is implicitly included, the SFH posterior will also be skewed
towards this sSFH peak depending on the relative amplitude and width of the
likelihood distribution. 
In other words, the inferred SFH will generally be flatter as a function of
$\tlb$ than the true SFH. 
We can see this effect in Figure~\ref{fig:sfh_demo}. 
For the star-forming galaxy with a relatively flat SFH at intermediate values,
the inferred SFH is in good agreement with the true SFH. 
However, for the quiescent galaxy, which has high SFRs at early times ($\tlb >
6$ Gyr) and low SFRs at late times ($\tlb < 2$ Gyr), the inferred SFH is
flatter and skewed towards intermediate values. 
We note that the prior on SFH is similar to the priors on SFHs by various
nonparametric SPS models in \cite{leja2019}. 
Any detailed analysis of SFHs (\emph{e.g.} quenching timescale or star
formation variable) based on SED modeling must take the impact of model
imposed
priors on SFH into account or taken with a grain of salt.  

We emphasize that all SPS models impose undesirable priors on derived galaxy
properties. 
And \emph{any} deviation of the priors on galaxy properties from a uniform
distribution impacts the posteriors on the galaxy properties. 
Galaxy properties derived from SED modeling must, therefore, characterize
and account for the priors imposed on them by the model for unbiased and
accurate analyses. 
In this appendix, we characterize the model imposed priors of our PROVABGS SED
model for the main galaxy properties that we explore in this work. 
This allows us to interpret the posteriors of galaxy properties for PROVABGS
and qualitatively disentangle the effect of the prior. 
Upcoming work in Hahn in prep. will demonstrate that maximum-entropy priors can
be used to substantially mitigate the impact of model impose priors on the
posteriors of galaxy properties (see Section~\ref{sec:discuss}). 

\section{Population Inference} \label{sec:hyper}
% restatement/high level explanation of what we do
We quantify the accuracy and precision of the inferred galaxy properties from
our SED modeling using population hyperparameters 
$\eta_{\Delta} = \{\mu_{\Delta_\theta}, \sigma_{\Delta_\theta}\}$ 
(Section~\ref{sec:results}). 
These hyperparameters describe the distribution of the difference between the
inferred and true parameters, $\Delta_{\theta}$, assuming that the distribution
has a Gaussian functional form (Eq.~\ref{eq:eta_gauss}). 
The $\eta_\Delta$ values we present in this work are MAP estimates of
$p(\eta_\Delta | \{ X_i \})$, the probability distribution of $\eta_\Delta$
given some galaxy population observations. 
They are inferred using population inference as described in the main text and
Eqs~\ref{eq:popinf} - \ref{eq:popinf2}.
Our approach for quantifying the accuracy and precision has a number of key
advantages over other methods. 
For instance, a naive way to quantify the accuracy and precision would be to
estimate the median and standard deviations of individual posteriors then
averaging them. 
This assumes that each individual posterior is close to a Gaussian. 
As we later demonstrate, this is an incorrect assumption that reduces the
posterior distribution to point estimates. 
Another approach would be to stack the posteriors by summing up all of the
individual posteriors. 
Neither of these approaches mathematically estimate the distribution we are
actually interested in estimating: $p(\eta_\Delta | \{ X_i \})$. 
Moreover, both approaches are biased. 
\cite{malz2020} recently demonstrated this in the context of combining
photometric redshift posteriors. 

\begin{figure}
\begin{center}
\includegraphics[width=0.45\textwidth]{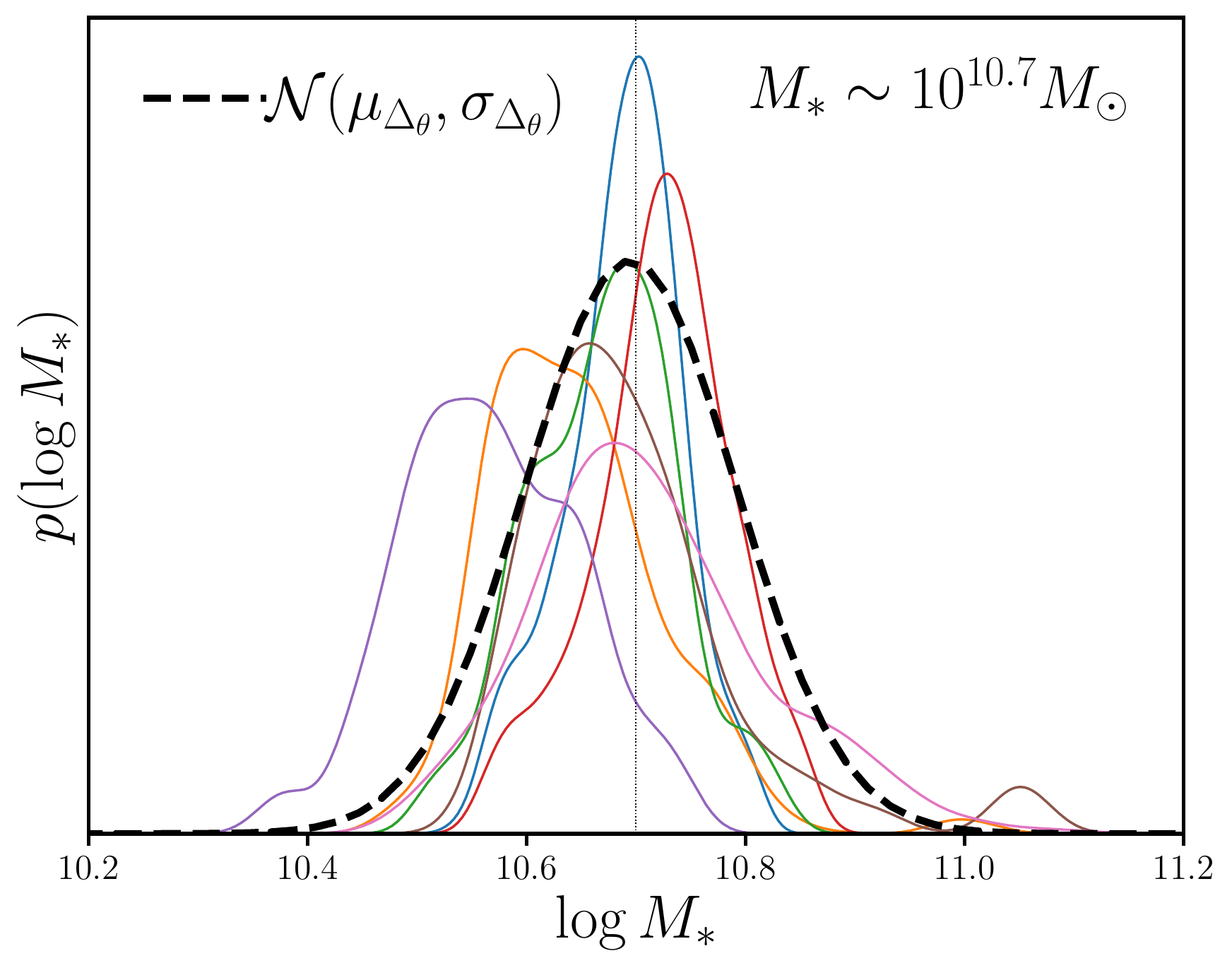}
    \caption{
    The $\log M_*$ distribution described by the accuracy and precision
    hyperparameters for galaxies with $10.6 < \log M_* < 10.8$:
    $\mathcal{N}(10.7 + \mu_{\Delta_\theta}, \sigma_{\Delta_\theta})$ 
    (black dashed).
    The hyperparameters are MAP estimates of $p( \mu_{\Delta_\theta}, 
    \sigma_{\Delta_\theta} | \{ X_i \})$ derived from population inference
    (Eq.~\ref{eq:popinf}-\ref{eq:popinf2}). 
    We include individual $\log M_*$ posteriors of several galaxies with $M_*
    \sim 10^{10.7} M_\odot$ for comparison.   
    The individual posteriors have a wide variety of shapes, which can bias 
    naive estimates of their accuracy and precision. 
    The comparison illustrates that 
    $\mathcal{N}(\mu_{\Delta_\theta}, \sigma_{\Delta_\theta})$ provides a
    robust estimate of the overall accuracy and precision of the inferred
    posteriors. 
    }\label{fig:eta_demo}
\end{center}
\end{figure}
We illustrate the population inference approach in Figure~\ref{fig:eta_demo} 
where we present the distribution of $\log M_*$ described by the accuracy and
precision hyperparameters derived for galaxies with $10.6 < \log M_* < 10.8$: 
$\mathcal{N}(\mu_{\Delta_\theta} + 10.7, \sigma_{\Delta_\theta})$ (black
dashed). 
For comparison, we plot posteriors of $\log M_*$ for several individual
galaxies with $\log M_* \sim 10.7$.  
There is significant variation in the individual posteriors and many of them
are not well described by a Gaussian distribution. 
This variation is an expected consequence of noise in the observables and
MCMC sampling.  
We note that estimating the accuracy and precision by stacking the posteriors,
for instance, significantly underestimates the precision.  
Meanwhile, the accuracy and precision hyperparameters capture the overall
accuracy and precision of the individual posteriors.

\bibliographystyle{mnras}
\bibliography{gqp_mc} 
%\allauthors
\end{document}